\documentclass{article}
\usepackage{arxiv}

\usepackage{authblk}
\usepackage{graphicx}
\usepackage{newtxtext}
\usepackage{bm}
\usepackage{newtxmath}
\usepackage{natbib}
\usepackage{hyperref}
\usepackage{mathtools}
\usepackage{placeins}
\hypersetup{
    colorlinks = true,
    urlcolor   = blue,
    citecolor  = black,
}

\newcommand{\RomanNumeralCaps}[1]
\linenumbers
\usepackage{booktabs} 
\usepackage{multirow} 
\newcommand*{\diff}{\mathop{}\!\mathrm{d}}

\title{Using Optimal Transport Aligned Latent Embeddings for Separated Flow Analysis}

\author[1]{Jonathan Quang Tran\thanks{Email: jqtranus@g.ucla.edu}}
\author[2]{Chi-An Yeh}
\author[1]{Kunihiko Taira}

\affil[1]{Department of Mechanical and Aerospace Engineering, University of California, Los Angeles, CA 90095, USA}
\affil[2]{Department of Mechanical and Aerospace Engineering, North Carolina State University, Raleigh, NC 27695, USA}

\begin{document}
\maketitle

\begin{abstract}
Quantifying differences between flow fields is a key challenge in fluid mechanics, particularly when evaluating the effectiveness of flow control or other problem parameters. Traditional vector metrics, such as the Euclidean distance, provide straightforward pointwise comparisons but can fail to distinguish distributional changes in flow fields. To address this limitation, we employ optimal transport (OT) theory, which is a mathematical framework built on probability and measure theory. By aligning Euclidean distances between flow fields in a latent space learned by an autoencoder with the corresponding OT geodesics, we seek to learn low-dimensional representations of flow fields that are interpretable from the perspective of unbalanced OT. { As a demonstration, we utilize this OT-based analysis on separated flows past a NACA 0012 airfoil with periodic heat flux actuation near the leading edge}. The cases considered are at a chord-based Reynolds number of 23,000 and a freestream Mach number of 0.3 for two angles of attack of $6^\circ$ and $9^\circ$. For each angle of attack, we identify a two-dimensional embedding that succinctly captures the different effective regimes of flow responses and control performance, characterized by the degree of suppression of the separation bubble and secondary effects from laminarization and trailing-edge separation. The interpretation of the latent representation was found to be consistent across the two angles of attack, suggesting that the OT-based latent encoding was capable of extracting physical relationships that are common across the different suites of cases. This study demonstrates the potential utility of optimal transport in the analysis and interpretation of complex flow fields.
\end{abstract}

\begin{keywords}
Authors should not enter keywords on the manuscript, as these must be chosen by the author during the online submission process and will then be added during the typesetting process (see \href{https://www.cambridge.org/core/journals/journal-of-fluid-mechanics/information/list-of-keywords}{Keyword PDF} for the full list).  Other classifications will be added at the same time.
\end{keywords}


\section{Introduction}
\label{sec:Introduction}

The analysis of flow fields and the design of effective control strategies require an exhaustive comparison of flows across typically large parameter spaces. This frequently involves substantial qualitative and quantitative analyses of a large number of high-fidelity simulations, experimental data, or a combination of both. Given the high-dimensional, nonlinear, and multi-scale nature of fluid flows, it is often advantageous to attempt to understand differences in key features of the flows through low-order models that extract relevant features ~\citep{Lumely1967, schmid2010dynamic, taira2017modal}.  

Recently, the use of autoencoder (or encoder-decoder) methods for dimensionality reduction has grown in popularity due to their ability to handle complex nonlinear problems as opposed to traditional linear techniques~\citep{murata2020nonlinear, vinuesa2022enhancing, zeng2022data, racca2023predicting, FT2023, smith2024cyclic, tran2024aerodynamics, mousavi2025low}. These models seek to approximate a continuously differentiable and invertible transformation between a manifold in some high-dimensional space and a subset of some low-dimensional latent space. \citet{murata2020nonlinear} demonstrated the use of a convolutional autoencoder to compress the dynamics of flow past a cylinder to nonlinear modes, exhibiting superior compression performance and robustness to noise compared to proper orthogonal decomposition (POD). ~\citet{zeng2022data} utilized a low-order space learned by an autoencoder to perform reinforcement learning for the control of chaotic systems in a tractable manner. Additionally,~\citet{tran2024aerodynamics} exhibited the effectiveness of a combined POD and neural network autoencoder approach for performing a data-driven aerodynamic design optimization of automobile geometries for drag reduction.

Although autoencoders are a powerful method to compress and extract features from fluid flows, they come with a few drawbacks. While methods such as POD may perform worse in compressing highly complex nonlinear systems, the low-order coordinates and modes have intuitive physical explanations and have reproducible results~\citep{taira2017modal}, which is not explicitly true for most machine-learning-based approaches. Fundamentally, autoencoders operate on the principle that high-dimensional datasets exist on low-dimensional manifolds~\citep{gorban2018blessing}. The autoencoder learns a latent space which gives curvilinear coordinates that describe a parametric surface fit to the data~\citep{arvanitidis2017latent, magri2022interpretability}. Often these models are prone to overfitting, requiring extensive regularization and hyperparameter tuning to improve their generalization capabilities, especially when the underlying manifold is not densely sampled~\citep{kingma2013auto, lee2022regularized,kvalheim2023should}. The problem of learning the manifold and latent space coordinates is inherently ill-posed as any solution is only unique up to a diffeomorphism~\citep{pmlr-v251-syrota24a}. The non-uniqueness of the learned latent space coordinates is a major contributing reason to why obtaining a consistent interpretation of the latent coordinates is difficult outside of qualitative observations. This is especially an issue in the context of fluid mechanics, where it is desirable to use the learned latent space to perform downstream tasks that rely on geometric properties of the latent coordinates, such as dynamics modeling, flow field comparison, generative modeling, or interpolation.

Previous studies have sought to address some of these issues of interpretability. Studies by \citet{magri2022interpretability} and \citet{kelshaw2024proper} interpret the latent space geometry of autoencoders using proper latent decomposition, which extends POD approaches to the non-Euclidean geometries learned by autoencoders. \citet{FT2023} utilizes an augmented autoencoder approach to analyze gust-airfoil interactions in which a model is simultaneously trained to reconstruct flow-field information and an estimate of the lift coefficient from the compressed latent representation. This encourages the model to learn a low-order representation where there is a qualitative relationship between the latent space coordinates and estimates of the lift. \citet{smith2024cyclic} employed persistent homology, a method from topological data analysis, to design an autoencoder that provides a simple representation of the topology of the state dynamics of large-amplitude gust encounters in the latent space representation. 

In this study, we seek an alternative data-driven approach to imposing structure in the low-order coordinates for the analysis of fluid flows by incorporating information of pairwise distances or similarities into the learned latent coordinates. This is motivated by the fact that when evaluating the effect of actuation on the dynamics of a flow, it is often necessary to quantify a degree of similarity or dissimilarity between two distinct flow fields. This typically involves the use of metrics or distances. One example is the widely adopted $L^2$ metric (or Euclidean, $\ell^2$, for finite-dimensional vectors). However, while simple to compute, in some scenarios the Euclidean distance may not be the most informative measurement of dissimilarity as it ignores the underlying structural and geometric information, only facilitating pointwise comparisons of the flow field across corresponding spatial locations. This may be misleading in some scenarios, such as in the case where we seek to compare fields in which structures have been displaced. 


Here, we utilize optimal transport (OT), which offers an intuitive framework for quantifying dissimilarities between flow fields. The class of OT distances, which includes the Wasserstein distance or earth mover's distance, has been studied extensively for various problems, including those in economics~\citep{vaserstein1969markov, kantorovich1960mathematical}, generative modeling~\citep{arjovsky2017wasserstein}, image science~\citep{rubner1998metric,peyre2019computational}, and partial differential equation theory~\citep{villani2009optimal,mainini2012well}. Rather than only considering pointwise differences, OT distances effectively quantify a cost of transporting some resource or measure from one distribution to another~\citep{villani2009optimal}. This makes OT particularly well-suited for scenarios in which the comparison of the spatial arrangement of flow features --- such as vortices, coherent structures, or transported quantities --- is of interest. Additionally, OT distances have been demonstrated to be robust to noise and outliers in measurements~\citep{peyre2019computational}.  

To demonstrate the potential of OT applied to fluid flows, we consider the analysis of separated aerodynamic flows. Separated flow is a phenomenon accompanied by often undesirable behavior, such as lift reduction and drag increase~\citep{lissaman1983low, mueller2003aerodynamics}. Accordingly, engineers have long sought to understand and mitigate or reduce separation in aerodynamic flows through active and passive flow control techniques~\citep{prandtl1925magnuseffekt,lachmann1961boundary,wu1998post,seifert1999oscillatory,greenblatt2000control,joslin2009fundamentals,kotapati2010nonlinear,yeh2019resolvent}. 

Flow separation typically occurs due to the emergence of an adverse pressure gradient, in which the boundary layer is decelerated and detaches from the surface, forming a shear layer~\citep{lachmann1961boundary,mueller2003aerodynamics, Marxen:JFM2013}. Separation can also appear in the presence of sharp gradients in the surface geometry~\citep{lachmann1961boundary,Haggmark:PTRSL2000} or when induced by nearby vortical structures~\citep{harvey1971flowfield}. 
In the case of laminar separation, the separated shear layer is unstable, resulting in flow unsteadiness and/or a transition to turbulent flow~\citep{Dovgal:PAS1994}. For high Reynolds number flows, the separated shear layer becomes more receptive to instabilities and becomes highly unsteady, eventually fully transitioning to turbulence around which the Reynolds number is on the order of $10^4-10^5$. In such scenarios, the Kelvin-Helmholtz instabilities in the shear layer lead to the roll-up of spanwise coherent vortical structures, which drive momentum-mixing between the free-stream and the reverse flow region~\citep{tani1964low, yarusevych2009vortex, kotapati2010nonlinear, yeh2019resolvent}. Unsteady mixing facilitates entrainment of high-momentum fluid from the free stream to enable the separated shear layer to reattach to the airfoil surface further downstream, provided the boundary layer is sufficiently energized~\citep{lachmann1961boundary,lissaman1983low, mueller2003aerodynamics,jaroslawski2023disturbance}. 

When the flow reattaches, the time-averaged flow exhibits a closed region of recirculating flow, forming a separation bubble. Depending on various factors, such as chord-based Reynolds number or the angle of attack, the characteristics of this separated flow can vary substantially. The size, shape, and position of the separation bubble directly influence the performance of the airfoil, including stall characteristics and lift-to-drag ratio~\citep{tani1964low,lissaman1983low, mueller2003aerodynamics,klewicki2025aerodynamic}. At Reynolds numbers on the order of $10^5$, we may observe a long separation bubble that occupies over 20-30\% of the chord, resulting in noticeable changes in the pressure distribution over the suction side~\citep{lissaman1983low}, while at even higher Reynolds numbers, a short separation bubble may be observed. These shorter bubbles are generally characterized by a linear increase of lift with the angle of attack, with the bubble bursting at the onset of stall~\citep{lissaman1983low,mueller2003aerodynamics}. However, a long separation bubble can also lead to stall without bursting if it spans a sufficient amount of the airfoil surface.

The unsteady separation bubble has been described as a self-excited flow structure maintained by a feedback loop that occurs due to interactions between the Kevin-Helmholtz instability in the shear layer and the vortex shedding in the wake~\citep{kiya1997sinusoidal, Zaman:JFM1989, greenblatt2000control, yarusevych2009vortex, kotapati2010nonlinear}. Various studies have been performed to investigate how to leverage these instability mechanisms that contribute to the sustainment of the separation bubble. In particular, periodic forcing has been demonstrated to be an effective method of unsteady boundary layer control, as it can reach similar control authority as steady blowing or suction while requiring orders of magnitude smaller momentum input~\citep{wygnanski1997boundary, greenblatt2000control}. \citet{amitay2002role} performed an experimental parametric study using surface-mounted fluidic actuators to observe the flow response over a range of forcing frequencies, finding that unsteady periodic actuation can support complete flow reattachment. \citet{yeh2019resolvent} explored the use of resolvent analysis to guide the design of a periodic thermal flux actuator for the suppression of flow separation over a NACA 0012 airfoil. The resolvent formulation was used to gain insights into the effects of variation of the spatial and temporal actuation frequencies of a heat flux actuator on the separation dynamics.

In what follows, we analyze the effect of unsteady thermal actuation on separated flow over a NACA 0012 airfoil in a data-driven manner from the perspective of OT. We leverage an OT distance-based embedding learned by an autoencoder to analyze the controlled flows in a low-dimensional manner. We align pairwise Euclidean distances in the latent space with dissimilarities computed with OT, which attempts to associate the geometric structure of the latent space with the spatial distribution of fluid structures according to the OT distance. This restriction on the latent space coordinates allows relative distances in the latent space to encode information of similarity in terms of the spatial distribution of structures in the flow response. We find that the distribution of flow responses to periodic thermal actuation is reducible to an interpretable two-dimensional latent representation that encodes the behavior of the separation bubble in the time-averaged flows. This demonstrates the potential utility of OT in the analysis of fluid flows.

\section{Problem Setup and Approach}
\label{sec:Approach}

\subsection{Optimal Transport for Comparison of Fluid Flows}
\label{sec:OT}

\begin{figure}
  \centerline{\includegraphics[width=1\textwidth]{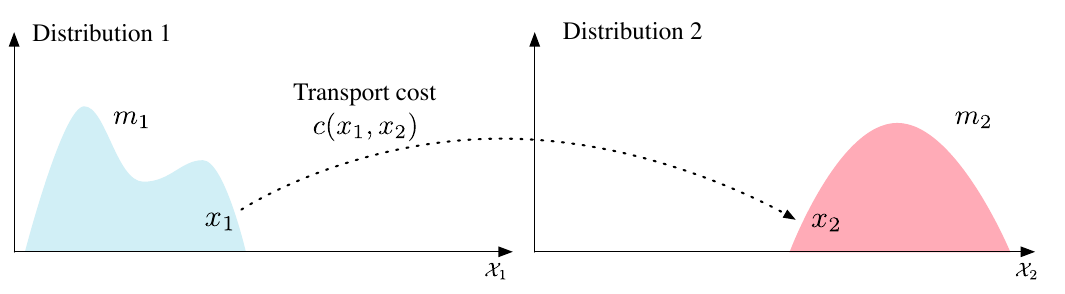}} 
  \caption{Pedagogical schematic of optimal transport between a supply and demand distribution representative of two different flow fields. The optimal transport distance is the minimum total cost to move the supply to the demand distribution.}
\label{fig:optimal_transport}
\end{figure}

We utilize OT distances to quantify the change in the distribution of structures in flow fields. In this section, we give a brief description of OT and its application in computing dissimilarities between flow fields. While our focus is on a high-level explanation, additional details can be found in the Appendix. For a more comprehensive review of the relevant subjects and formalisms from probability theory and optimal transport, we refer the reader to~\citet{williams1991probability},~\citet{villani2009optimal}, and~\citet{tao2011introduction}.

OT is naturally framed within the language of probability theory and measures. To define a measure, we first start with the concept of a \textit{measurable space}, which comprises of a set $\mathcal{X}$ and a $\sigma$-algebra on $\mathcal{X}$ denoted as $\mathcal{A}$, which is a nonempty collection of subsets of $\mathcal{X}$ that is closed under complementation, countable unions, and countable intersections. 
A measure is then a set function that takes measurable sets from $\mathcal{A}$ and maps them to an element of the extended real numbers. This measurement must satisfy certain properties, such as being zero for an empty set and also being countably additive. A physical interpretation could be the measurement of the mass, which prescribes a numerical value (mass) to a physical region of space.

Loosely speaking, the distribution of a measure describes how it is organized over the measurable space $(\mathcal{X},\mathcal{A})$. For example, if $\mathcal{X}$ represents a physical domain, a distribution can describe the spatial arrangement of measures like the density of a material, the concentration of a substance, or the probability of an event occurring in a region. In our discussion of fluid flows, we can consider distributions of physical measures of interest, where the measure corresponds to a quantity such as the fluid density or velocity magnitude. 

OT distances provide a mathematical framework for comparing distributions of quantities while accounting for spatial structure. Originally developed to address problems of efficient resource allocation, the original optimal transport problem seeks the most cost-effective way to redistribute a given resource from one configuration to another~\citep{vaserstein1969markov, kantorovich1960mathematical}. The classical formulation considers a set of supply (or source) locations over which resources are distributed, and a set of demand (or target) locations, each requesting a specified amount of resources. The cost of transportation between locations depends on both the ground cost, in this case, the distance traveled, and the quantity of resources transported. The OT distance then corresponds to the minimum cost required to transport the resources from the supply distribution to the demand distribution. For these cases in which the total supply and demand are the same (balanced), the source and target distributions can be represented as probability measures, and OT can be used to define a metric between them. An illustration of OT for this pedagogical problem is shown in figure~\ref{fig:optimal_transport}(a). 

Following these examples, one can intuitively understand OT distances as a standard way to lift metrics (or distances) defined on some ground measurable space to a distance between distributions of measures~\citep{villani2009optimal, peyre2019computational, chizat2016scaling}. OT enables us to leverage the underlying geometric structure to compare distributions. {In the context of fluid flows, the compared distributions may represent different flow fields, with each distribution describing the spatial variation of physical quantities such as density, velocity, or vorticity.}

{ To define the OT distance between two measures, we start with spaces corresponding to the spatial domain of the two distributions, which are depicted in figure~\ref{fig:optimal_transport}(a) by $\mathcal{X}_1$ and $\mathcal{X}_2$. 
We then choose a lower semi-continuous cost function that maps between the product space of $\mathcal{X}_1$ and $\mathcal{X}_2$ to the non-negative extended reals $C : \mathcal{X}_1 \times \mathcal{X}_2 \to \mathbb{R}_{\geq 0}\cup\{+\infty\}$. This cost function $C$ describes the cost of transport from point $x_1 \in \mathcal{X}_1$ to another point $x_2 \in \mathcal{X}_2$. For this work, we choose $C$ to be the Euclidean distance between points in $\mathcal{X}_1$ and $\mathcal{X}_2$. However, the relevant choice of the cost function is problem dependent. We note that the inclusion of $+\infty$ in the range is to account for possible obstacles that may be present in the flow domain (e.g., a solid submerged body). Given $\mathcal{M}_+( \cdot )$, the set of all nonnegative measures over some measurable space, we have that for positive measures $m_1 \in \mathcal{M}_+(\mathcal{X}_1)$ and $m_2 \in \mathcal{M}_+(\mathcal{X}_2)$, the optimal transport (OT) cost is defined as:
\begin{align}
\mathcal{J}(\Gamma) \coloneqq \int_{\mathcal{X}_1 \times \mathcal{X}_2} C(x_1,x_2) \, \mathrm{d}\Gamma(x_1,x_2).
\end{align}
With this cost, the balanced OT distance is computed with the following constrained minimization:
\begin{equation}
d_\text{OT}(m_1, m_2) \coloneqq 
\underset{\Gamma \in \mathcal{M}_+(\mathcal{X}_1 \times \mathcal{X}_2)}{\inf}
\left\{
\mathcal{J}(\Gamma)
:
P^{\mathcal{X}_1}_{\#}\Gamma = m_1, P^{\mathcal{X}_2}_{\#}\Gamma = m_2
\right\}.
\end{equation}
Here $\Gamma$, called the coupling (or transport plan), is a nonnegative measure over the product space $\mathcal{X}_1 \times \mathcal{X}_2$ and describes the transport of resources between the distributions of $m_1$ and $m_2$. The joint distribution of $\Gamma$ over $\mathcal{X}_1 \times \mathcal{X}_2$ quantifies how much resource is moved between the points $x_1$ and $x_2$. The measures $P^{\mathcal{X}_1}_{\#}\Gamma$ and $P^{\mathcal{X}_2}_{\#}\Gamma$ denote the first and second marginals of $\Gamma$, respectively. These are defined identically to the familiar notion of marginal probability, given by $P^{\mathcal{X}_1}_{\#}\Gamma = \int_{x_2\in\mathcal{X}_2}d\Gamma(\, \cdot \, , x_2)$ and $P^{\mathcal{X}_2}_{\#}\Gamma = \int_{x_1\in\mathcal{X}_1}d\Gamma(x_1, \, \cdot \,)$. The constraints enforce that the total measure must be the same for both distributions.}


Naturally, the distributions of measures in different fluid flow fields need not integrate to the same amount. However, the traditional balanced OT formulation requires both distributions to have the same total measure. To account for this imbalance, we utilize a modification to the OT cost using Csisz\'{a}r divergence functionals. These divergences are functionals that compare two distributions of measures in a pointwise manner (like the Kullback-Leibler divergence) and can be formally built with entropy functions in the context of probability and information theory. {Further details on how these divergences are computed for the general measures in the current work can be found in Appendix~\ref{appA}.}
{ The total cost of $\mathcal{J}(\Gamma)$ is modified to include an extra divergence functional term, replacing the equality constraint on the marginals in the optimization. In other words, instead of strictly enforcing equality of the total measures, the divergences only softly penalize imbalance between the distributions, giving us an unbalanced OT distance.}


\begin{figure}
  \centerline{\includegraphics[width=1\textwidth]{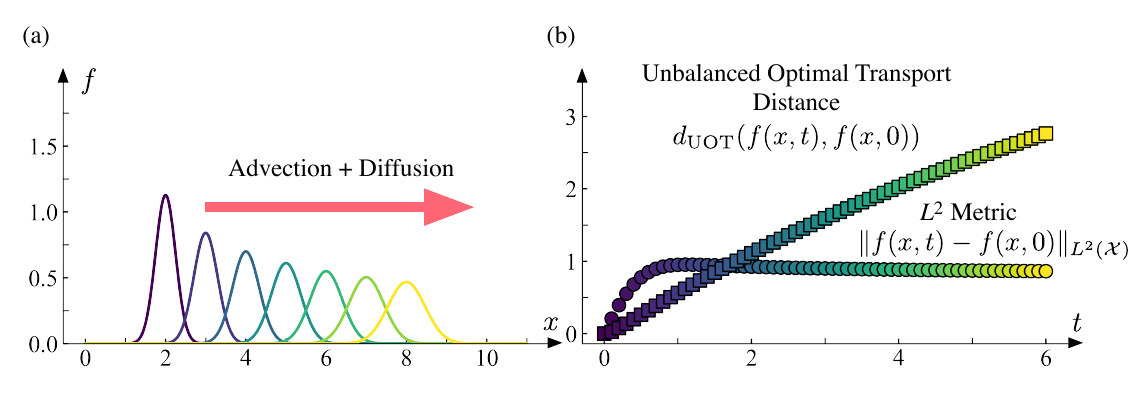}} 
  \caption{Comparison of unbalanced optimal transport distance $d_{\text{UOT}}$ with the $L^2$ metric. (a) Gaussian pulse undergoing advection and diffusion. (b) Comparison of distances between the evolving pulse and the initial condition.}
\label{fig:optimal_transport_advection}
\end{figure}

Thus, to allow for a possible imbalance between the two flow fields, we additionally consider $\mathcal{D}_{\varphi_1}$ and $\mathcal{D}_{\varphi_2}$, which are Csisz\'{a}r divergences over $\mathcal{X}_1$ and $\mathcal{X}_2$. Given $\mathcal{M}_+( \cdot )$, the set of all nonnegative measures over some measurable space, we have that for positive measures $m_1 \in \mathcal{M}_+(\mathcal{X}_1)$ and $m_2 \in \mathcal{M}_+(\mathcal{X}_2)$, the unbalanced optimal transport (UOT) cost is defined as:
\begin{align}
\mathcal{J}(\Gamma) \coloneqq \int_{\mathcal{X}_1 \times \mathcal{X}_2} C(x_1,x_2) \, \mathrm{d}\Gamma(x_1,x_2)
+ \rho\left[\mathcal{D}_{\varphi_1}(P^{\mathcal{X}_1}_{\#}\Gamma | m_1) 
+ \mathcal{D}_{\varphi_2}(P^{\mathcal{X}_2}_{\#}\Gamma | m_2)\right].
\end{align}
Here, the term $\rho>0$ is described as a characteristic radius of transport which balances the contribution of the transport and divergence terms~\citep{sejourne2019sinkhorn}. The UOT distance is then defined as the infimum of this cost over $\Gamma \in \mathcal{M}_+(\mathcal{X}_1 \times \mathcal{X}_2)$:
\begin{equation}
d_\text{UOT}(m_1,m_2) \coloneqq \underset{\Gamma \in \mathcal{M}_+(\mathcal{X}_1 \times \mathcal{X}_2)}{\inf} \mathcal{J}(\Gamma).
\end{equation}
The first integral term describes the cost associated with a change in the spatial distribution. For example, this term would capture if a flow structure, such as a vortex or a shear layer, is at a different position in two flow fields. The second term that includes the divergences penalizes how different the marginals of $\Gamma$ are from the input measures $m_1$ and $m_2$, which allows for imbalance in the total amount of the two input measures. 

To compute the divergence functionals, we use the Kullback-Leibler entropy and we set the characteristic transport radius $\rho = 1$~\citep{villani2009optimal, peyre2019computational, chizat2016scaling}. {. To solve the optimization problem, we use the Sinkhorn–Knopp algorithm~\citep{chizat2016scaling}. The Sinkhorn–Knopp algorithm is an iterative matrix-scaling procedure that efficiently computes an entropy-regularized optimal transport plan. It alternates between normalizing the rows and columns of a transport matrix until convergence, making it well-suited for large-scale problems, as it only requires scaling operations instead of solving the full linear system. This yields an approximation $S^\varepsilon(m_1,m_2) \approx d_\text{UOT}(m_1,m_2)$. We utilize the Python Optimal Transport library, which efficiently implements the Sinkhorn-Knopp matrix scaling algorithm~\citep{flamary2021pot}.

For two flow fields $V_1$ and $V_2$ we define corresponding measures $m_1$ and $m_2$, corresponding to the distribution of some physical variable. To allow for the case of signed measures, such as vorticity, we compute the UOT distance for the positive and negative parts of the flow field separately and sum to obtain a total distance as follows
\begin{equation}
d_\text{field}(V_1,V_2) = S^\varepsilon(m_1^+,m_2^+) + S^\varepsilon(m_1^-,m_2^-).
\end{equation}
To consider multiple measures corresponding to the same field (e.g., distributions of multiple components of velocity), we can compute a total dissimilarity $\tilde{d}_{\text{field}}(\mathbf{V}_1,\mathbf{V}_2)$ by some permutation invariant aggregation of the dissimilarities for each variable $d_\text{field}(V_1^{(i)},V_2^{(i)})$. A detailed explanation of how the flow field dissimilarity $\tilde{d}_{\text{field}}(\mathbf{V}_1,\mathbf{V}_2)$ is computed can be found in Appendix~\ref{appB}.
}

To build further physical intuition, consider figure~\ref{fig:optimal_transport_advection}(a), which shows selected time steps of the solution to the advection-diffusion equation, with a Gaussian initial condition. Presented in figure~\ref{fig:optimal_transport_advection}(b) are the $L^2$ metric and the unbalanced OT distance $d_{\text{UOT}}$ from the initial condition. { Note that for this particular example, one can also opt to use the balanced OT distance, given that the shown transport process is conservative}. Here, by $L^2$ metric, we mean the metric that is induced by the $L^2$ norm on the Lebesgue space of square integrable functions on some domain $\mathcal{X}$ (as opposed to the $\ell^2$ metric for vectors in $\mathbb{R}^n$). The $L^2$ distance from the initial condition is given by  
\begin{equation}
    \| f(\cdot, t) - f(\cdot, 0) \|_{L^2(\mathcal{X})} = \left\{ \int_{\mathcal{X}} \left[ f(x, t) - f(x, 0) \right]^2 \, \mathrm{d}x \right\}^{1/2},
\end{equation}
which captures pointwise dissimilarities. As a result, it increases rapidly with even just a small translation but then saturates, classifying most subsequent time steps equally dissimilar once there is little overlap with the initial condition. Unlike the $L^2$ metric, the unbalanced OT distance correlates with the displacement of the Gaussian pulse even when the support does not overlap with the initial condition. As a result, the unbalanced OT distance provides a characterization of dissimilarity that better aligns with our intuitive understanding of how the distribution evolves over time.

\subsection{Physical Problem Description}
\label{sec:setup}


As a demonstrative example, we consider the effects of active flow control on separated flows over a NACA 0012 airfoil at angles of attack $\alpha \in \{ 6^{\circ} , 9^{\circ}\}$ with chord-based Reynolds number $Re_{L_c} \equiv u_{\infty}L_c/\nu_\infty=23,000$ and free stream Mach number $M_{\infty} \equiv  u_{\infty}/a_{\infty} = 0.3$. Here $u_\infty$ is the free-stream velocity, $L_c$ is the chord length, $a_\infty$ is the free-stream speed of sound, and $\nu_\infty$ is the kinematic viscosity. 

The flow fields were obtained via large eddy simulation (LES) using the $CharLES$ finite-volume compressible flow solver ~\citep{khalighi2011unstructured,bres2017unstructured} with the Vreman subgrid model~\citep{vreman2004eddy}. We used a C-grid mesh following the setup in~\citet{yeh2019resolvent} which has been examined for convergence in the flow field and aerodynamic forces with refinements the near-field. The computational domain covers $(x/L_c,y/L_c,z/L_c)\in[-19,26]\times[-20,20]\times[-0.1,0.1]$ with the leading edge of the airfoil being placed at the origin. The solution was computed using a constant timestep of $\Delta t u_\infty / L_c = 4.14 \times 10^{-5}$ corresponding to a maximum Courant-Friedrichs-Lewy (CFL) number of 0.84. The statistics of the aerodynamic forces were computed using over 80 convective time units. Additional details of the computational setup and grid are provided in~\citet{yeh2017use} and~\citet{yeh2019resolvent}. 

The specific heat ratio was specified as $\gamma = 1.4$ with a Prandtl number $Pr \equiv \nu /\kappa = 0.7$, where $\kappa$ is the thermal diffusivity. The dynamic viscosity was specified as a function of temperature in the form of a power law, given by $\mu(T) = \mu_\infty(T/T_\infty)^{0.76}$ for $T/T_\infty \in [0.5,1.7]$ with $\mu_\infty$ and $T_\infty$ denoting the free stream dynamic viscosity and temperature, respectively~\citep{garnier2009large}. For all of the flows in this study, we observed that the maximum temperature fluctuation in the flow is within 42\% of $T_\infty$, falling within the allowable range of the viscosity model~\citep{yeh2019resolvent}.

For the far-field boundary, a Dirichlet boundary condition was specified as $(\rho,u_x,u_y,u_z,T)=(\rho_\infty,u_\infty,0,0,T_\infty)$. A sponge layer~\citep{freund1997proposed} was placed along the outlet boundary with a target state being the running-averaged flow over 10 acoustic time units. Over the airfoil surface, a no-slip boundary condition was prescribed. The airfoil surface was also specified to be adiabatic except where the periodic heat-flux actuator was placed at the leading edge.

\begin{figure}
  \centerline{\includegraphics[width=0.75\textwidth]{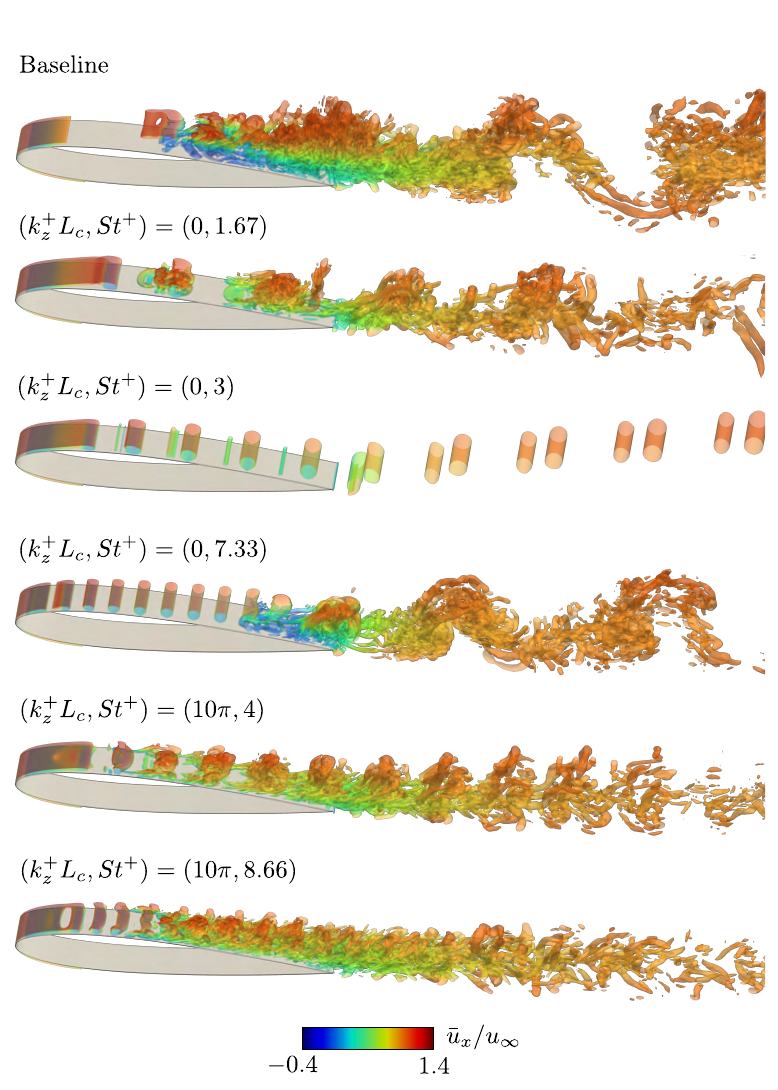}} 
  \caption{Examples of baseline flow and various responses of a separated wake past a NACA 0012 airfoil at $\alpha=6^\circ$ to a heat flux actuator input at the leading edge \citep{yeh2019resolvent}. The $Q L_c^2/u_\infty^2=50$ isosurface colored by the normalized streamwise velocity is shown.}
\label{fig:case_example}
\end{figure}

The thermal actuation setup employed in this study followed the methodology outlined in~\citet{yeh2019resolvent}, where a thermal actuator was placed across the span near the leading edge with prescribed frequency and spanwise profile. The actuator was implemented in the energy equation as a nonhomogeneous, time-dependent Neumann boundary condition. The actuator model was expressed using a Hann window with compact spatial support given by
\begin{equation}
    \phi^{+}(\omega^{+}, k_z^{+}) = \frac{1}{4} \hat{\phi} \sin(\omega^{+} t) \left[ 1 + \cos(k_z^{+} z) \right] \left\{ 1 + \cos \left[ \frac{2\pi}{\sigma_a}(x - x_a) \right] \right\},
\end{equation}
where $(x-x_a)/\sigma_a \in [-0.5,0.5]$. The actuator was centered at $x_a/L_c = 0.03$ on the suction surface with width $\sigma_a/L_c=0.04$. The amplitude $\hat{\phi}$ was fixed according to the normalized total actuation power,
\begin{equation}
    E^+ = \frac{\frac{1}{4}\hat{\phi}\sigma_a}{\frac{1}{2}\rho_\infty v_\infty^3 (L_c \sin\alpha)}=0.0902,
\end{equation}
which is consistent with the setup of previous studies~\citep{corke2010dielectric,sinha2012impulse,yeh2019resolvent}. The thermal input from the actuator manifests in an oscillatory surface vorticity flux and baroclinic torque~\citep{yeh2017laminar,yeh2019resolvent}.

From this actuator, we have a two-dimensional parameter space of control inputs $(k_z^+,\omega^+)\in 10\pi\mathbb{Z} \times \mathbb{R}_{\ge 0}$. The control parameter pairs are characterized by a dimensionless wavenumber $k_z^+L_c$ and chord-based actuation Strouhal number $St^+ \coloneq \omega^+ L_c/(2 \pi u_\infty)$. For each angle of attack, we considered actuation wavenumbers of $k_z^+L_c \in \{0, 10\pi,20\pi,40\pi\}$ with around 30 cases varying $St^+ \in [0, 18]$ for a total of around 120 cases per angle of attack. 

An assortment of wake responses resulting from changes in the input parameter \( St^+ \) are shown in figure~\ref{fig:case_example}. Despite varying only two parameters in the shown cases, the resulting flow fields exhibit remarkably complex and diverse behavior. These differences manifest in several key aspects of the wake dynamics. For example, the extent and nature of flow separation deviates greatly, with the size and position of the separation bubble differing significantly across different controlled cases. Additionally, the degree of laminarization and turbulent mixing over the airfoil surface and in the downstream wake also changes noticeably between cases, with some cases exhibiting partial or global laminarization. The wide array of flow field responses motivates the use of a data-driven analysis to extract relevant features in the flow response that relate to the control performance.

\subsection{OT Dissimilarity Based Coordinate Identification}
\label{sec:autoencoder}

\begin{figure}
  \centerline{\includegraphics{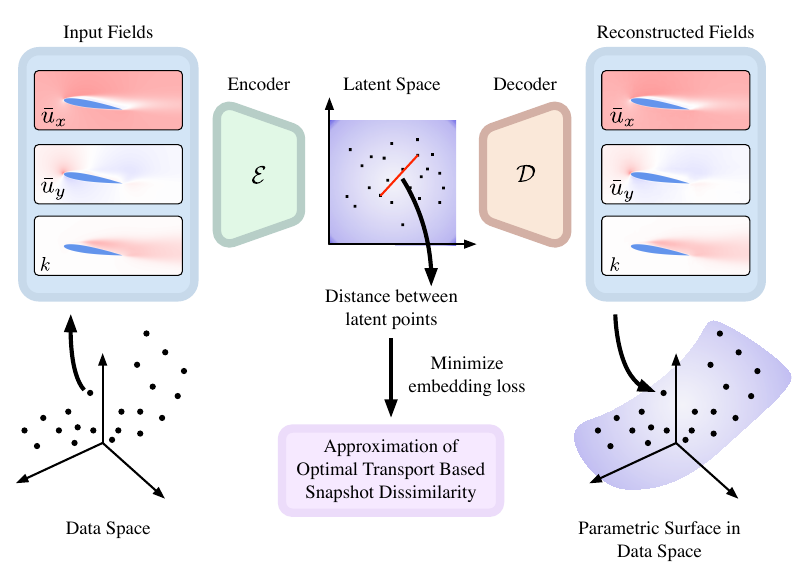}} 
  \caption{Illustration of the augmented autoencoder problem setup. During training, the model learns to associate Euclidean distances in the latent space (red line) with flow field dissimilarities computed with optimal transport.}
\label{fig:MDS_AE}
\end{figure}

We seek to utilize a modified autoencoder to learn a low-dimensional representation of fluid flows while preserving dissimilarity between inputs in the learned latent representation. In other words, the goal is to learn not just a compressed representation of our flow fields, but to learn a representation with some geometric structure informed by OT that allows the latent coordinates to be interpreted as providing relative similarities between flow fields. 

Consider some instance of discretized flowfield data $\mathbf{q} \in \mathbb{R}^{N \times N_s}$ where $N$ refers to the number of variables and $N_s$ refers to the size of the spatial discretization. A typical autoencoder consists of an encoder-decoder pair $(\mathcal{E},\mathcal{D})$ which can be constructed via neural networks ~\citep{HS2006}. The encoder $\mathcal{E} : \mathbb{R}^{N \times N_s} \to \mathbb{R}^n$ maps the input from a physical space to a latent space representation, with coordinates $\bm{\xi} \in \mathbb{R}^n$. The decoder $\mathcal{D} : \mathbb{R}^n \to \mathbb{R}^{N \times N_s}$ takes a latent space coordinate and produces a reconstruction of the input $\tilde{\mathbf{q}} = \mathcal{D} \circ \mathcal{E}(\mathbf{q}) \in \mathbb{R}^{N \times N_s}$. In this study, the encoder network consists of convolutional layers followed by dense layers. Each layer is followed by batch normalization and a ReLU activation function. The decoder is constructed in a reverse fashion with dense layers followed by convolutional layers. Residual skip connections are placed around each convolutional block to stabilize training and reduce overfitting~\citep{he2016deep,ioffe2015batch}. { The parameters for the autoencoder architecture are given in Table~\ref{tab:nn_architecture}. In the present work, the number of filters are increased with depth~\citep{lecun2002gradient}. All variables have been min-max normalized prior to training.

\begin{table}
  \begin{center}
  \def~{\hphantom{0}}
  \begin{tabular}{ll|ll}
    \toprule
    \multicolumn{2}{c}{\textbf{Encoder}} & 
    \multicolumn{2}{c}{\textbf{Decoder}} \\
    \midrule
    Layer & Data Size & 
    Layer & Data Size \\
    \midrule
    Input $(\bar{u}_{x}, \bar{u}_{y},k)$ & (250, 100, 3) & Dense & (64) \\
    Conv2D + BN (3,3,16) & (250, 100, 16) & Dense & (128) \\
    MaxPooling2D & (125, 50, 16) & Dense & (256) \\
    Conv2D + BN (3,3,32) & (125, 50, 32) & Dense & (1536) \\
    MaxPooling2D & (63, 25, 32) & Reshape & (8, 3, 64) \\
    Conv2D + BN (3,3,32) & (63, 25, 32) & Conv2D + BN (3,3,32) & (8, 3, 32) \\
    MaxPooling2D & (32, 13, 32) & UpSampling2D & (16, 6, 32) \\
    Conv2D + BN (3,3,64) & (32, 13, 64) & Conv2D + BN (3,3,32) & (16, 6, 32) \\
    MaxPooling2D & (16, 6, 64) & Conv2D + BN (3,3,32) & (16, 6, 32)\\
    Conv2D + BN (3,3,64) & (16, 6, 64) & UpSampling2D & (32, 12, 32) \\
    MaxPooling2D & (8, 3, 64) & Conv2D + BN (3,3,16) & (32, 12, 16) \\
    Conv2D (3,3,64) & (8, 3, 64) & Conv2D + BN (3,3,16) & (32, 12, 16) \\
    Reshape & (1536) & UpSampling2D & (64, 24, 16) \\
    Dense & (256) & Conv2D + BN (3,3,16) & (64, 24, 16) \\
    Dense & (128) & Conv2D + BN (3,3,16) & (64, 24, 16) \\
    Dense & (64) & UpSampling2D & (128, 48, 16) \\
    Dense (Latent vector) & (2) & Conv2D + BN (3,3,8) & (128, 48, 8) \\
    - & - & Conv2D + BN (3,3,8) & (128, 48, 8) \\
    - & - & UpSampling2D & (250, 100, 8) \\
    - & - & Output $(\bar{u}_{x}, \bar{u}_{y},k)$ & (250, 100, 3) \\
    \bottomrule
  \end{tabular}
  \caption{Network architecture of the encoder and decoder models. The activation function used is ReLU.}
  \label{tab:nn_architecture}
  \end{center}
\end{table}}

Here, we consider the time and spanwise averaged streamwise velocity, vertical velocity, and the turbulent kinetic energy, $k$, as the input and output, i.e., $\mathbf{q} = (\bar{u}_{x}, \bar{u}_{y},k)$. 
We note that since $\bar{\omega}_z = \partial \bar{u}_y/\partial x - \partial \bar{u}_x/\partial y$, we can also reconstruct $\bar{\omega}_z$ using the reconstructed velocities for assessment of the reconstruction performance. For the autoencoder input and output as well as the flow field OT-dissimilarity computation, we consider a two-dimensional spatial region around the airfoil given by $\mathcal{X} = [-0.5L_c,2L_c] \times [-0.5L_c,0.5L_c]$ with $n_x=250$ and $n_y=100$. 

The model's weights $\theta$ are optimized via the following objective function:
\begin{align}
    \theta^* &= \underset{\theta}{\text{argmin}} \ \ (\mathcal{L}_{1} + \lambda \mathcal{L}_{2}) \\
    \mathcal{L}_{1} &= \frac{1}{M N N_s}\sum_{i=1}^M \left\lVert  \mathbf{q}_i- \mathcal{D} \circ \mathcal{E}(\mathbf{q}_i) \right\rVert_F^2 \\
    \mathcal{L}_{2} &= \frac{2}{M(M-1)}\sum_{i=1}^M\sum_{j=i+1}^M\left( \lVert  \mathcal{E}(\mathbf{q}_i)-\mathcal{E}(\mathbf{q}_j)\rVert_2 -  \tilde{d}_{\text{field}}(\mathbf{q}_i,\mathbf{q}_j) \right)^2,
\end{align}

where $M$ is the dataset size. The first term in the objective $\mathcal{L}_1$ is the mean square error of the flow field reconstruction, which seeks to make the autoencoder approximate the identity function such that $\mathcal{D} \approx \mathcal{E}^{-1}$. 

\begin{figure}
  \centerline{\includegraphics[width=1\textwidth]{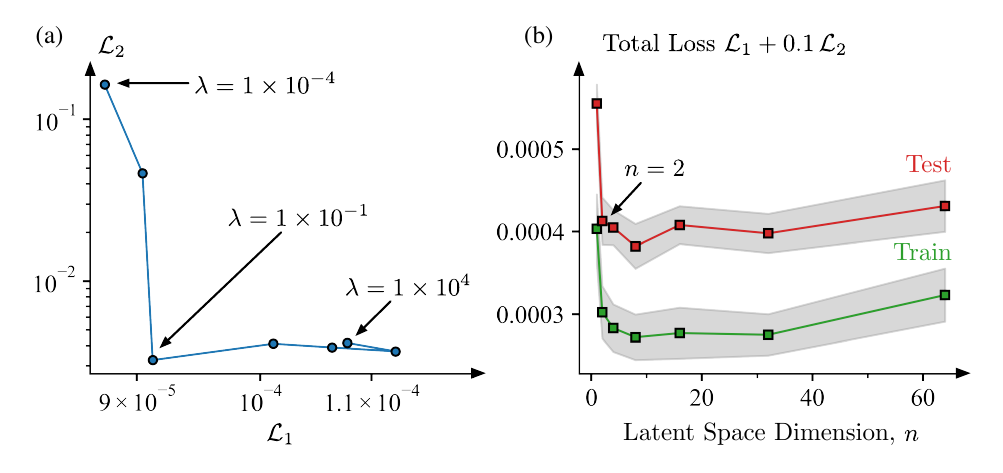}} 
  \caption{Example parameter study of the OT-based autoencoder for $\alpha=9^\circ$. (a) L-curve showing the trade-off between $\mathcal{L}_1$ and $\mathcal{L}_2$ for the test set with $10^{-4} \le \lambda \le 10^4$. (b) Variation of the total loss $\mathcal{L}_1 + \lambda \mathcal{L}_2$ with respect to the latent space dimension for $\lambda = 0.1$. Standard deviation for the last 500 epochs is colored in gray.}
\label{fig:param_study}
\end{figure}

We introduce the OT-based embedding loss term $\mathcal{L}_2$, defined as the squared difference comparing pairwise Euclidean distances of latent points with the OT-based flow field dissimilarities of their corresponding flow fields. With this term, we seek to impose a geometric structure on the embedded flow fields in which straight lines in the latent space should correspond to an approximate OT geodesic between inputs. If structures in the flow field are moved between two different inputs, resulting in a higher flow field dissimilarity, the model would essentially learn to embed these fields further apart in the latent space. Conversely, if two inputs have a small flow field dissimilarity, the model should learn to embed them close to one another in the latent space. This prevents the autoencoder from placing latent points arbitrarily relative to one another. With such an embedding, geometric properties of the latent space can be used to intuit physical trends or relationships between different controlled cases. We note that if necessary for downstream tasks, we can also include a secondary decoder $\mathcal{F} : \mathbb{R}^n \to \mathbb{R}^m$ which maps from latent coordinates to estimates of physical parameters~\citep{FT2023, tran2024aerodynamics}. However, for our particular demonstration, we found that using the relevant performance metrics $\mathbf{p} = (\bar{C}_{D},\bar{C}_{L}) \in \mathbb{R}^2$ as an auxiliary output did not have a significant impact on the overall results.

We utilize the AdamW optimizer~\citep{kingma2014adam} with an initial learning rate of $1 \times 10^{-4}$ and weight decay parameter of $10^{-2}$ for up to $4000$ total epochs. An $80:20$ split is performed between training and test data. The hyperparameter $\lambda$ determines the relative influence of the embedding loss term. We choose this parameter by observing the tradeoff between the two loss terms as we vary $\lambda$. The tradeoff takes the form of an $L$-curve, or Pareto front, and we choose $\lambda$ to be around the elbow of this curve~\citep{hansen1993use}.

\section{Results}

\begin{figure}
  \centerline{\includegraphics[width=0.75\textwidth]{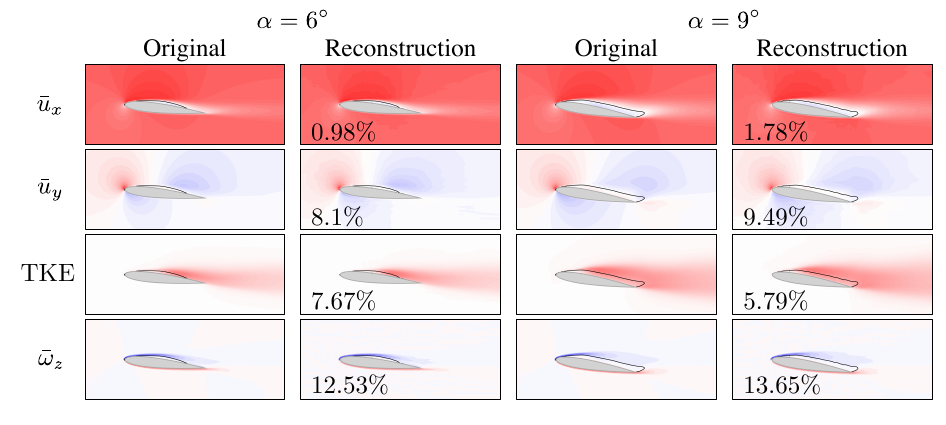}} 
  \caption{Reconstructions of baseline flow fields by OT autoencoder. The $\bar{u}_x = 0$ isocontour is shown in black for all fields. The reconstructed vorticity is obtained by central differencing of the reconstructed velocity fields. The percent Frobenius norm reconstruction error is reported.}
\label{fig:Reconstruction_Baseline}
\end{figure}

We examine the effectiveness of the present OT-based approach in identifying low-dimensional coordinates that succinctly capture the effects of the open-loop thermal control on the separated flow for controlled cases with $\alpha=6^\circ$ and $9^\circ$. In the following analyses, we consider separate autoencoders trained only on each angle of attack. For the choice of the $\mathcal{L}_2$ loss weight $\lambda$, an L-curve analysis is shown in figure~\ref{fig:param_study}(a) with $10^{-4} \le \lambda \le 10^4$. We choose $\lambda$ to be the value around the elbow of the curve, which occurs around when $\lambda=0.1$. In figure~\ref{fig:param_study}(b), we report the total loss for both training and test sets with respect to the choice of the latent space dimension with the chosen weight of $\lambda=0.1$. The mean and standard deviation of the loss are computed using the last 500 epochs of training, where the loss has stabilized. For both angles of attack, we find that the loss sharply drops as we go from $n=1$ to $n=2$, however, the loss for the test set seems to vary little after this point. We show a representative reconstruction from the OT autoencoder using a two-dimensional latent space for both angles of attack in figure~\ref{fig:Reconstruction_Baseline}. We report the reconstruction error for a given flow field image $f$ with $\varepsilon=100 \times \lVert f_{\text{rec.}}-f_{\text{original}}\rVert_F/\lVert f_{\text{original}}\rVert_F$ for each reconstructed field. To additionally assess the autoencoder performance, we compare $\omega_z$ computed from the input and reconstructed velocities. We observe that the OT autoencoder is capable of qualitatively reconstructing the flow field from just a two-dimensional latent space, successfully recovering the general profile of the separation bubble and shear layer.

{
The performance of the two models (with $n=2$) for both angles of attack is summaried in table~\ref{tab:err_combined}. The average errors for the flow fields $f$ are reported using the Frobenius norm error metric.
Note that if the discretized flow fields are flattened from matrices into vectors, this is equivalent to the vector $\ell_2$ norm, which is usually reported. For the OT embedding term, error is measured using a normalized $\ell_2$ error metric between the embedded latent distances and the OT-based dissimilarity of the corresponding flow fields:
\begin{equation}
    \varepsilon_{d}= 100 \times \sqrt{\frac{\sum_{i=1}^M\sum_{j=i+1}^M\left( \lVert  \mathcal{E}(\mathbf{q}_i)-\mathcal{E}(\mathbf{q}_j)\rVert_2 -  \tilde{d}_{\text{field}}(\mathbf{q}_i,\mathbf{q}_j) \right)^2}{\sum_{i=1}^M\sum_{j=i+1}^M \lVert  \mathcal{E}(\mathbf{q}_i)-\mathcal{E}(\mathbf{q}_j)\rVert_2^2}},
\end{equation}
which in this context is referred to as the multi-dimensional scaling (MDS) stress~\citep{kruskal1978multidimensional,davison2000multidimensional,borg2005modern}. This is a measurement of how distorted the embedded distances are from the original dissimilarities, and is invariant under translation and uniform stretching of the dissimilarities and latent coordinates. An MDS stress value that is smaller than $15\%$ is typically described as acceptable, although this threshold is heuristic and can be problem dependent~\citep{kruskal1964nonmetric,kruskal1978multidimensional,borg2005modern}.
Additional details of the training performance for the training and test sets are reported in Appendix~\ref{appC}.

\begin{table}
  \begin{center}
  \def~{\hphantom{0}}
  \begin{tabular}{llcccccc}
    \toprule
    AoA & Quantity         & \multicolumn{2}{c}{Reg. AE Err.}  & \multicolumn{2}{c}{OT AE Err.} \\
         &                 & Train & Test & Train & Test \\
    \midrule
    \multirow{6}{*}{$6^\circ$}
         & $\bar{u}_x$      & 1.11  & 1.13  &  1.25  & 1.37 \\
         & $\bar{u}_y$      & 9.99  & 10.08 &  8.75  & 9.05 \\
         & $k$
         & 7.86  & 9.35  &  8.45  & 10.36 \\
         & $\bar{\omega}_z$ & 13.26 & 14.36 &  15.09 & 15.29 \\
         & OT Emb. & -     & -     &  12.77  & 13.52 \\
    \midrule
    \multirow{6}{*}{$9^\circ$}
         & $\bar{u}_x$      & 1.38  & 1.14  &  1.42  & 1.47 \\
         & $\bar{u}_y$      & 8.59  & 8.50  &  8.50  & 8.48 \\
         & $k$
         & 6.62  & 7.18  &  8.28  & 8.71 \\
         & $\bar{\omega}_z$ & 16.60 & 16.48 &  14.89 & 14.76 \\
         & OT Emb.     & -     & -     &  13.82  & 15.00 \\
    \bottomrule
  \end{tabular}
  \caption{Comparison of average percent error for field variables using different autoencoder architectures (with two latent variables) at angles of attack $\alpha = 6^{\circ}$ and $9^{\circ}$. Field variable errors are reported as percent Frobenius norm error. The embedding loss is reported using the MDS stress.}
  \label{tab:err_combined}
  \end{center}
\end{table}

\FloatBarrier
}

\subsection{$\alpha = 9^\circ$ Latent Space}
\begin{figure}
  \centerline{\includegraphics[width=1\textwidth]{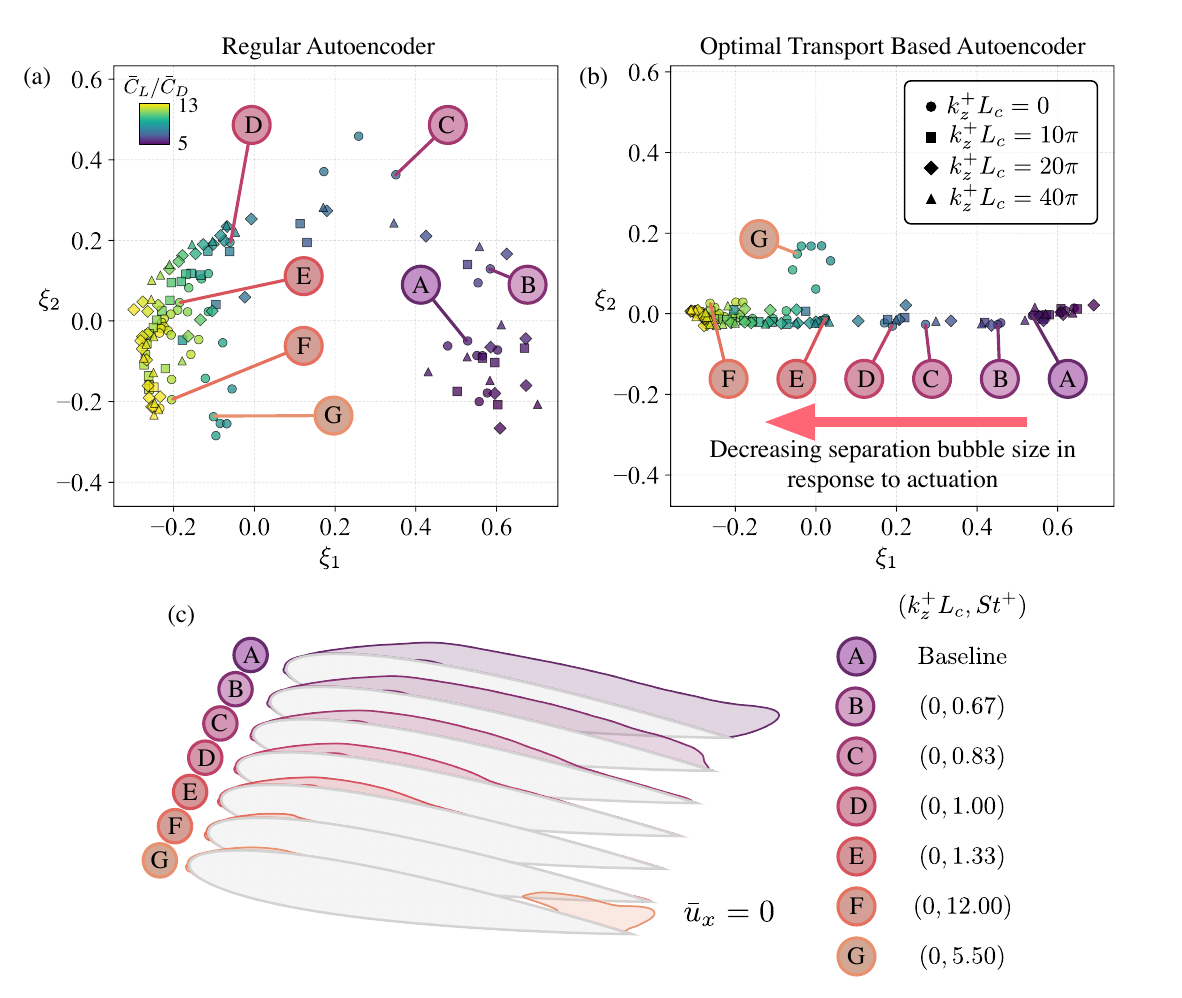}} 
  \caption{Comparison of learned latent spaces with $\alpha=9^\circ$ colored by aerodynamic performance $\bar{C}_L/\bar{C}_D$ for (a) standard autoencoder ($\mathcal{L}_1$ loss only) (b) OT autoencoder ($\mathcal{L}_1$ and $\mathcal{L}_2$ losses). (c) Isocontours of time-averaged streamwise velocity $\bar{u}_x = 0$ shown for different representative cases labeled in (a) and (b).}
\label{fig:AoA9_Compare_Lat}
\end{figure}

We begin by examining the $\alpha = 9^\circ$ cases, which exhibit comparatively simpler flow responses than the $\alpha = 6^\circ$ cases, primarily due to the absence of global laminarization in the flow responses~\citep{yeh2019resolvent}. In particular, we first observe the effect of the OT based embedding term on the learned latent space representation. We consider latent spaces learned by two representative autoencoder models as shown in figure~\ref{fig:AoA9_Compare_Lat}: (a) a standard autoencoder trained solely with reconstruction loss ($\mathcal{L}_1$) and (b) the OT autoencoder which includes the OT embedding term ($\mathcal{L}_1$ and $\mathcal{L}_2$). For ease of discussion, we rotate all latent spaces to align with their principal component axes so that we can describe how the flow fields vary along axes where the OT-distances are estimated to vary the most. This transformation does not affect the loss or the interpretation of relative dissimilarity as the reconstruction loss $\mathcal{L}_1$ remains unchanged since the rotation can be reversed without loss, and the $\mathcal{L}_2$ term is inherently invariant to uniform rotation of the latent space. We note that both models achieve comparable reconstruction performance, indicating that the additional loss term does not severely compromise the ability to accurately recover the flow fields. 

\begin{figure}
  \centerline{\includegraphics[width=0.9\textwidth]{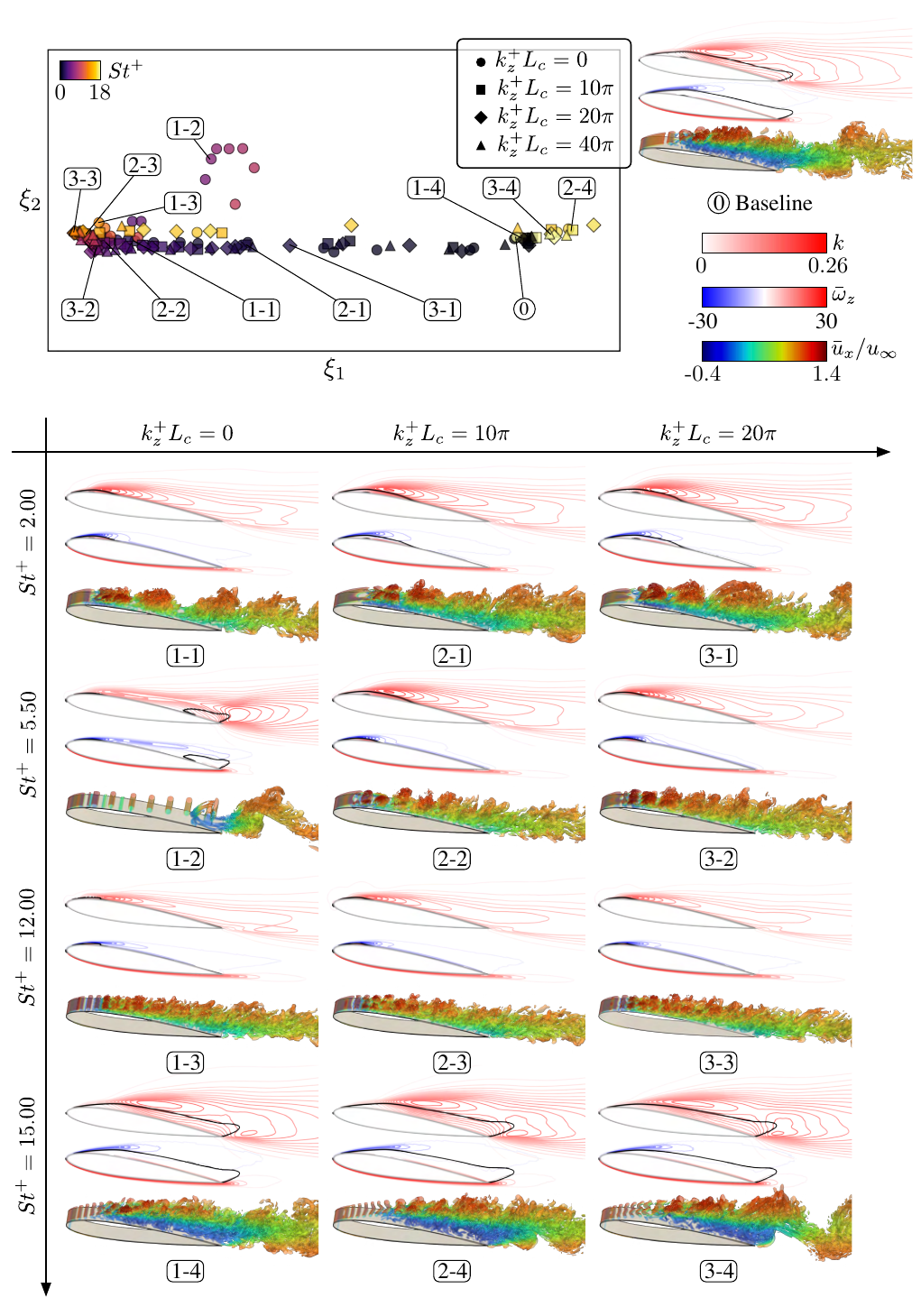}} 
  \caption{Plot of latent embeddings highlighting the influence of actuation parameters for $\alpha=9^\circ$. Labeled are example actuation cases. Examples of average turbulent kinetic energy, average vorticity fields, as well as instantaneous $Q$-criterion ($Q L_c^2/u_\infty^2 = 50$) colored by streamwise velocity are shown for the labeled cases. The $\bar{u}_x = 0$ isocontour is shown for all average fields.}
\label{fig:AoA9_Compare_Kz}
\end{figure}

Also depicted in figure~\ref{fig:AoA9_Compare_Lat} are the $\bar{u}_x=0$ isocontours for representative input cases, which exhibit the variation in the separation bubble behavior as we traverse the latent space. Case A in figure~\ref{fig:AoA9_Compare_Lat} shows the baseline flow, with a long separation bubble that spans the entire suction side. As we move from case A to F, we observe that the separation bubble shrinks towards the leading edge. Conversely, case G presents a qualitatively different flow regime. Separation occurs near the trailing edge as spanwise rollers merge and break down near the trailing edge, forming a recirculation zone, and there is a pronounced increase in turbulent kinetic energy near the trailing edge. This is more clearly shown by case 1-2 in figure~\ref{fig:AoA9_Compare_Kz}.

For both the regular autoencoder and the OT based latent space, we see a sequential progression from case A to F as the separation bubble shrinks, which correlates with the control performance in terms of the average lift-to-drag ratio. However, for the standard autoencoder in figure~\ref{fig:AoA9_Compare_Lat}(a), while we can make out a qualitative trend of the lift-to-drag ratio in the latent space, the latent representations are unconstrained in geometry, and there is no inherent interpretation to the distance between latent points. We again emphasize that the relative positioning of the latent points for the standard autoencoder latent space in figure~\ref{fig:AoA9_Compare_Lat}(a) is arbitrary up to diffeomorphism, which inhibits our ability to judge how similar the flow fields are based solely on their latent positions. In fact, the distinction between the outlier cases in which separation occurs near the trailing edge (case G) is not clearly made in the regular autoencoder latent space. Additionally, in the latent space learned by the regular autoencoder, the best-performing and worst-performing cases begin to approach each other, despite being the most qualitatively ``dissimilar'' cases.

In the OT-based latent space shown in figure~\ref{fig:AoA9_Compare_Lat}(b), we find that the first latent coordinate $\xi_1$ clearly corresponds with the variation in size of the separation bubble. As $\xi_1$ decreases, the separation bubble size reduces and eventually vanishes, as seen in cases A-F. The second latent coordinate $\xi_2$ captures changes in the flow field associated with the partial laminarization of the flow and the trailing-edge separation, which is specific to the $k_z^+=0$ configurations, depicted by case G. Most cases are distributed along the $\xi_1$ axis (i.e., first principal component), suggesting that the separation bubble size is the dominant mode of variation in the flow response according to the OT-based latent geometry. We once again emphasize that the ability to derive an interpretation for the principal components in the latent space arises because Euclidean distances in the latent space serve as surrogates for the OT-based snapshot dissimilarity.

From figure~\ref{fig:AoA9_Compare_Lat}(b), we additionally observed that $\xi_1$ is also strongly correlated with the control performance in terms of the average lift-to-drag ratio. Lower values of $\xi_1$ are associated with reduced flow separation (hence a smaller separation bubble) and an increased lift-to-drag ratio. To understand the variation of the control performance in terms of the control input parameters, in figure~\ref{fig:AoA9_Compare_Kz} we show the OT-based latent space colored by the actuation frequency with marker shapes corresponding to the spanwise wave number of actuation. We find that the $\xi_1$ coordinate or separation bubble size primarily depends on $St^+$, having a weaker dependence on $k_z^+ > 0$.

We also depict several representative cases from both two and three-dimensional actuation at various frequencies in figure~\ref{fig:AoA9_Compare_Kz}, with their corresponding latent embeddings labeled. Starting from the baseline (case 0), increasing the forcing frequency within the range $0 \lesssim St^+ \lesssim 4.5$ (e.g., cases 1-1, 2-1, and 3-1) leads to a rapid decrease of $\xi_1$, corresponding to an improved lift-to-drag ratio as was seen in figure~\ref{fig:AoA9_Compare_Lat}(b). In this regime, flow separation is effectively suppressed, and the turbulent kinetic energy in the wake becomes lower, reflecting a reduction in the size of the separation bubble and the associated turbulent mixing region. We see that the two-dimensional actuation initially sees a faster rate of reduction of the separation bubble size with respect to increasing the actuation frequency, as seen in both the flow fields. This is also reflected in the relative placement of cases 1-1, 2-1, and 3-1 along $\xi_1$.

For $5 \lesssim St^+ \lesssim 8$, when $k_z^+ = 0$, there is an increase of $\xi_1$ and $\xi_2$ toward the cluster of outlier points (represented by case 1-2). Here, performance deteriorates for the two-dimensional actuation due to the reappearance of separation near the trailing edge, leading to a reduction lift-to-drag ratio (increase of $\xi_1$). In this range, two-dimensional actuation induces partial laminarization over the suction surface, characterized by spanwise vortical structures that originate near the leading edge and break down further downstream, causing the trailing edge separation. For $k_z^+>0$, trailing edge separation is not observed, and there is no substantial increase of $\xi_1$ during this intermediate range of forcing frequencies. We see that the OT-based embedding clearly distinguishes the $k_z^+ = 0$ cases with partial laminarization from the other control cases.

As the forcing frequency increases further to $9 \lesssim St^+ \lesssim 13$, the lift-to-drag ratio continues to improve, with the trailing edge separation suppressed for $k_z^+=0$ (case 1-3), and we return to the left side of the $\xi_1$ axis in latent space. It is during this range of forcing frequencies that we are the furthest in the negative $\xi_1$ direction, and we observe the local optimal lift-to-drag ratio. This occurs at around $St^+\approx12$ for $k_z^+=10\pi$ and $20\pi$ (cases 2-3 and 3-3). For $St^+ \gtrsim 12$, $\xi_1$ increases sharply, indicating a loss of actuator effectiveness. As a result, flow separation reappears, and lift-to-drag ratio declines, returning to baseline levels (cases 1-4, 2-4, 3-4).

\subsection{$\alpha = 6^\circ$ Latent Space}

\begin{figure}
  \centerline{\includegraphics[width=1\textwidth]{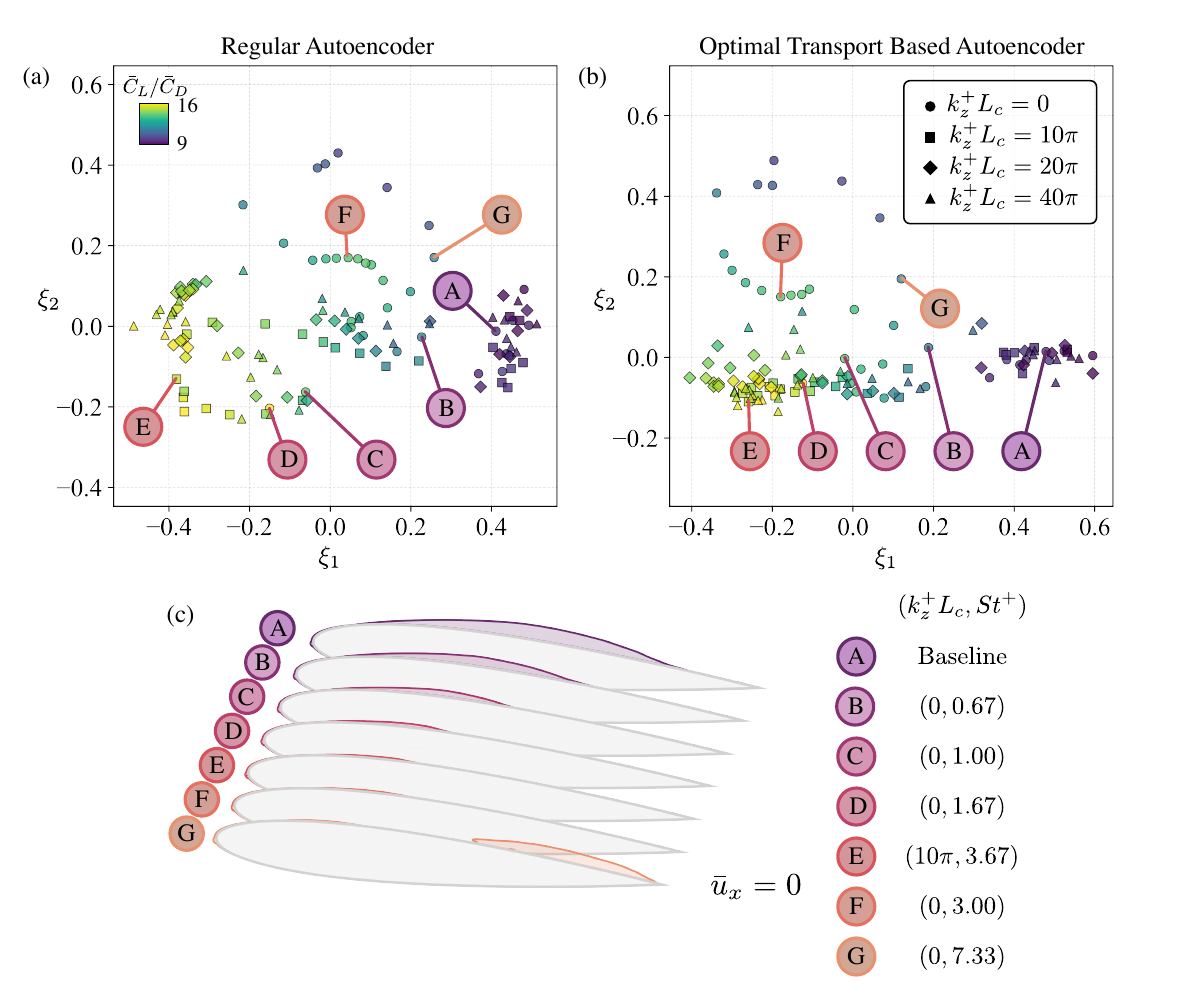}} 
  \caption{Comparison of learned latent spaces with $\alpha=6^\circ$ colored by control performance $\bar{C}_L/\bar{C}_D$ for (a) standard autoencoder ($\mathcal{L}_1$ loss only) (b) OT autoencoder ($\mathcal{L}_1$ and $\mathcal{L}_2$ losses). (c) Isocontours of time-averaged streamwise velocity $\bar{u}_x = 0$ shown for different representative cases labeled in (a) and (b).}
\label{fig:AoA6_Compare_Lat}
\end{figure}

Next, we analyze the latent embedding for the $\alpha = 6^\circ$ cases which exhibits a more diverse range of behaviors compared to $\alpha = 9^\circ$. To visualize the effect of the OT embedding for $\alpha=6^\circ$, in figure~\ref{fig:AoA6_Compare_Lat} we show the latent embeddings for the $\alpha=6^\circ$ cases corresponding to the regular and OT-based autoencoders in addition to separation profiles for representative cases. Just as in the $\alpha=9^\circ$ case, while see that we can interpret a trend of the lift-to-drag ratio for the regular autoencoder in figure~\ref{fig:AoA6_Compare_Lat}(a), we again emphasize that the relative placement of cases is arbitrary, and thus the cases are scattered in the latent space without any regard to any notion of similarity between the flow fields. We see that the OT-based embedding shown in figure~\ref{fig:AoA6_Compare_Lat}(b) more succinctly captures the separation bubble behavior (cases A-E) while distinguishing cases with laminarization and trailing edge separation (e.g., the branches containing cases F and G). In contrast to the $\alpha=9^\circ$ latent space, while we see that the lift-to-drag ratio visibly correlates with $\xi_1$, we also note that there is some dependence on $\xi_2$ as seen with the performance of some of these laminarized cases.
 
Again, as shown in figure~\ref{fig:AoA6_Compare_Lat}(b), the $\xi_1$-coordinate captures the dominant mode of variation in the latent space with most cases being positioned along this axis. As was the case for the $\alpha=9^\circ$ cases, a decrease in $\xi_1$ generally corresponds to a shrinking bubble that moves toward the leading edge, before disappearing completely. Likewise, the $\xi_2$-coordinate identifies control cases in which the flow exhibits different behavior in the wake due to laminarization. 
However, a major distinguishing feature of the $\alpha=6^\circ$ latent space compared to the $\alpha=9^\circ$ one is that for the cases with leading edge separation, the separation bubble size alone does not as effectively characterize distinct cases. For example, cases E and F have very similar separation bubble profiles; however, they clearly exhibit different control performance and placement in the latent space. To understand why this is the case, we note that unlike the $\alpha = 9^\circ$ cases, which only experience partial laminarization, the $\alpha=6^\circ$ latent space includes cases in which separation is suppressed and the flow field is globally laminarized (case F). To make this point clearer, in figure~\ref{fig:AoA9_Compare_Kz}, we again show the variation of the latent space with respect to the control parameters in addition to example average and instantaneous flow fields corresponding to labeled cases. For cases with global laminarization, spanwise vortical structures that originate from the leading edge indicate the suppression of three-dimensional flow structures as they advect downstream. These cases are distinguished from partially laminarized fields that have trailing-edge separation due to the breakdown of these spanwise structures (e.g., case G in figure~\ref{fig:AoA6_Compare_Lat}(c) or case 1-3 in figure~\ref{fig:AoA9_Compare_Kz}). This manifests in cases with very similar separation profiles, but different wake behavior.  
Additional effects associated with the laminarization of the flow captured in the $\xi_2$ coordinate are seen in the variation of the turbulent kinetic energy fields, which also play a significant role in the estimation of the lift-to-drag ratio for $\alpha=6^\circ$. 

\begin{figure}
  \centerline{\includegraphics[width=0.9\textwidth]{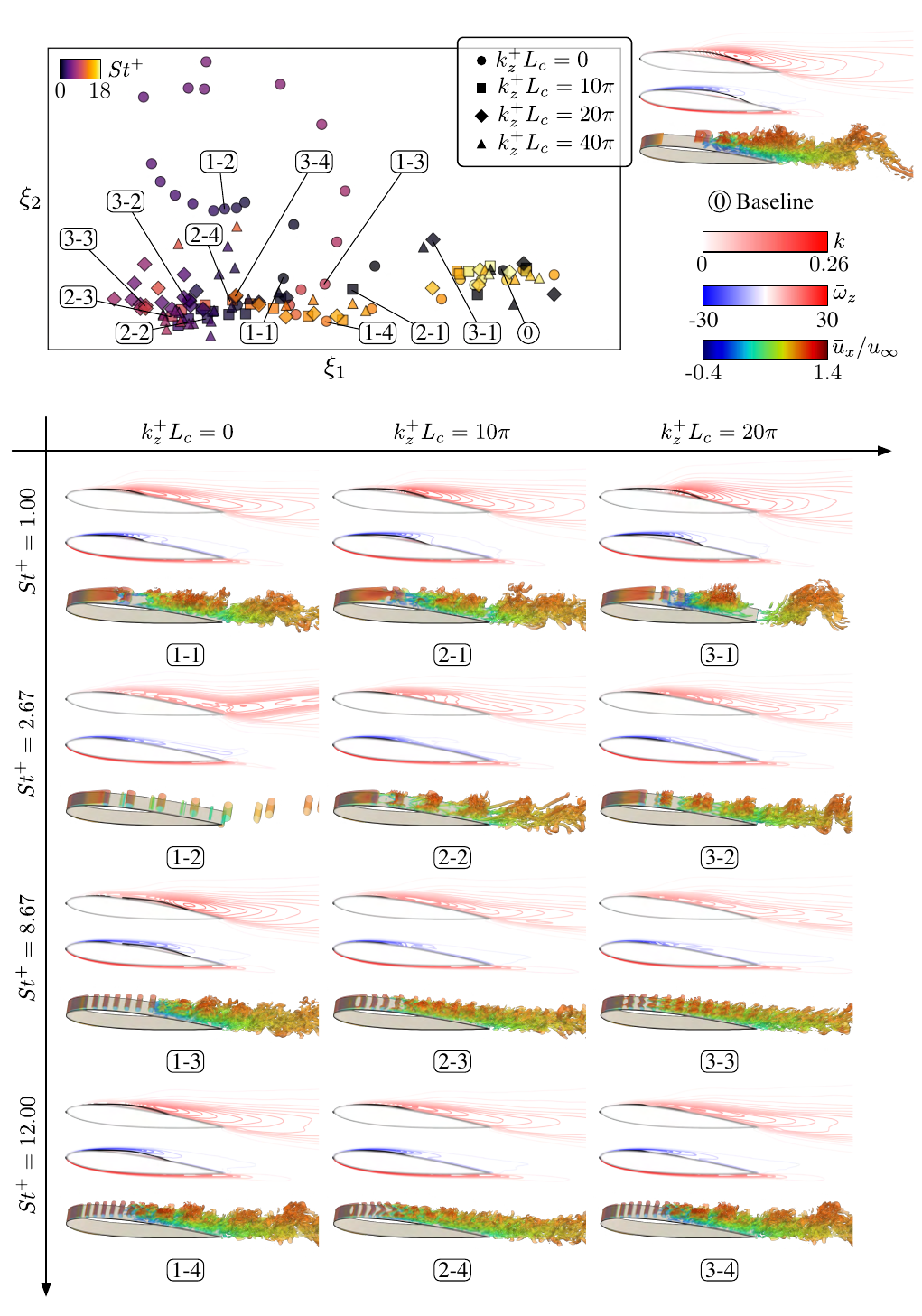}} 
  \caption{Plot of latent embeddings highlighting the influence of actuation parameters for $\alpha=6^\circ$. Labeled are example actuation cases. Examples of average turbulent kinetic energy, average vorticity fields, as well as instantaneous $Q$-criterion ($Q L_c^2/u_\infty^2 = 50$) colored by streamwise velocity are shown for the labeled cases. The $\bar{u}_x = 0$ isocontour is shown for all average fields.}
\label{fig:AoA6_Compare_Kz}
\end{figure}

Despite the aforementioned differences, the overall behavior within the latent space when the forcing frequency is varied parallels that of the $\alpha=9^\circ$ cases. If we start at the baseline flow (case 0) in figure~\ref{fig:AoA6_Compare_Kz}, increasing the forcing frequency within $0 \lesssim St^+ \lesssim 2$ results in an initial decrease of $\xi_1$ corresponding to the suppression of separation and improvement of the control performance (e.g., cases 1-1, 2-1, 3-1). 

For cases with $k_z^+ > 0$, increasing $St^+$ past this range would result in moving further left along $\xi_1$ until the local maxima in the lift-to-drag ratio is reached in the range of $2 \lesssim St^+ \lesssim 12$ as separation is eventually completely suppressed (e.g., cases 2-2, 3-2). For $k_z^+=0$, while $\xi_1$ is decreased, reflecting the suppression of separation, around $St^+ \approx 2$ ($\xi_1 \approx -0.1$), there is a vertical jump in $\xi_2$ corresponding to cases experiencing global laminarization (case 1-2). Here we see that the turbulent kinetic energy field shows a distinctively different appearance in the separated flow. The wake is directed upwards and is elongated towards the trailing edge, reflecting the advection of the pairs of spanwise rollers downstream. In this globally laminarized regime, even though there is a reattachment of the wake, there is a decrease in the lift-to-drag ratio when compared to cases with similar separation bubble sizes (similar $\xi_1$ position). 

When $k_z^+=0$ and $4 \lesssim St^+ \lesssim 8$, the vortical structures merge and break down near the trailing edge (e.g., case G in figure~\ref{fig:AoA6_Compare_Lat}), resulting in separation and a further decrease in performance. Eventually, the separation bubble begins to move back up towards the leading edge and rejoins the main distribution of cases along the $\xi_1$ axis (case 1-3). Moving between cases 1-2 and 1-3, we see that $\xi_2$ primarily characterizes differences in the trailing edge behavior. These cases observed strengthened turbulent kinetic energy mixing near the trailing edge and the wake, while cases along the $\xi_1$ axis primarily have the strongest turbulent kinetic energy at the rear edge of the separation bubble over the mid-chord and leading edge. For $St^+ \gtrsim 12$, $\xi_1$ increases for all $k_z^+$ as we again return to baseline performance as the controller has diminished effectiveness (e.g., cases 1-4, 2-4, 3-4).

In summary, these findings suggest that with the use of OT, the wide array of complex flow responses to thermal actuation can be effectively characterized by a low-dimensional representation that is consistent across the two angles of attack studied. For both $\alpha=6^\circ$ and $9^\circ$, the dominant mode of variation in response to control $\xi_1$ captures variations in separation bubble size and shows a strong correlation with the lift-to-drag ratio. Secondary effects are included in $\xi_2$, which encodes the behavior of the turbulent kinetic energy wake response corresponding to laminarization of the flow and the position of separation.

\section{Conclusion}


We introduced an autoencoder framework that incorporates unbalanced OT distances to learn latent representations of flow fields that encode a physically interpretable notion of similarity. Unlike standard autoencoders, which may organize latent variables arbitrarily, provided reconstruction accuracy is obtained, the OT-based approach imposes a notion of geometric structure by attempting to align pairwise distances in latent space with OT geodesics in the input space. When applied to flows over a NACA 0012 airfoil subject to unsteady thermal actuation, we are capable of obtaining an interpretable representation of the wide array of responses to control in a low-dimensional space. The use of OT as a dissimilarity metric in this problem setting allows us to capture displacements of flow structures and aerodynamic performance trends in response to different control inputs.

The current results demonstrate that the flow response to thermal actuation can be succinctly represented by two latent variables with the use of the OT-based embedding. The first latent coordinate captures the primary response of the flow to control. Along this coordinate, we explicitly observed a correlation between the separation bubble size and the control performance in terms of the lift-to-drag ratio. The second coordinate was found to contain information on changes in the wake due to laminarization of the flow and trailing edge separation. This interpretation of the learned latent coordinates in the OT-based approach was consistent across the two angles of attack studied. This consistency was not observed with the standard autoencoder formulation, suggesting the OT-based autoencoder uncovered shared physical relationships across the two sets of cases despite being trained separately for each angle of attack.

While this study employed a relatively simple scheme based on matching Euclidean distances in the latent space to OT-based dissimilarities in the input space, one may incorporate the intrinsic non-Euclidean geometry of the learned manifold. Although our analysis focused on a two-dimensional control parameter space, $(k_z^+, St^+)$, the present framework can be applied for higher-dimensional parameter spaces where interpretability becomes increasingly difficult and a naive exploration of parameters is expensive. Additionally, while the current study analyzed time-averaged flow fields, an extension to time-resolved or subsampled spatial domains~\citep{fukami2024single} presents a promising direction for the utilization of the OT-based approach for analyzing flow structures and control mechanisms. Additionally, it may be of interest to consider OT distances between state trajectories, rather than individual flow fields, using ground cost metrics such as the Hausdorff distance between attractors~\citep{ishar2019metric}, the Fr{\'e}chet distance \citep{alt1995computing}, or dynamic time warping~\citep{berndt1994using}. Given that we have also learned a relationship between the latent coordinates and the control performance, future efforts can also address design optimization in addition to uncertainty quantification. { Additionally, the present approach may assist in other downstream tasks such as generative modeling, given that interpolants in the latent space are trained to coincide with OT-geodesics, which may improve the quality of interpolation.}
With this in mind, the current results demonstrate the utility of OT in the representation and interpretation of complicated fluid physics.

\section*{Acknowledgments}{J.T. and K.T. acknowledge support from the U.S. Air Force Office of Scientific Research (FA9550-23-1-0715) and the U.S. Department of Defense Vannevar Bush Faculty Fellowship (N00014-22-1-2798). The authors thank Dr. Lionel Mathelin for insightful discussions on optimal transport.}


\section*{Declaration of interests}{The authors report no conflict of interest.}





\appendix
{
\section{Definition of Csisz\'{a}r divergence functionals}\label{appA}

Here we provide a high-level summary of Csisz\'{a}r divergence functionals as described in~\citet{chizat2016scaling},~\citet{,liero2018optimal}, and~\citet{nguyen2025introduction}. To compute a Csisz\'{a}r divergence functional, one must start with an entropy function. An admissible entropy function is a function $\varphi : \mathbb{R} \to \mathbb{R}_{\ge 0} \cup \{\infty\}$ which is convex, lower semicontinuous, and $\text{dom}(\varphi) \cap (0,\infty) \neq \emptyset$. 

From the entropy function, the Csisz\'{a}r divergence functional can be defined for two measures $m_1, m_2 \in \mathcal{M}(\mathcal{X})$ as:
\begin{equation}
\mathcal{D}_{\varphi}(m_1 \mid m_2) \coloneq 
\int_{\mathcal{X}} \varphi\!\left( \frac{\mathrm{d}m_1}{\mathrm{d}m_2} \right) \mathrm{d}m_2 
+ \varphi'_{\infty} m_1^{\perp}(\mathcal{X}),
\end{equation}
when $m_1$ and $m_2$ are nonnegative ($\infty$ otherwise). This expression relies on the Lebesgue decomposition, which decomposes $m_1$ into a part absolutely continuous with respect to $m_2$, denoted $m_1^{\mathrm{ac}}$, and a singular part $m_1^{\perp}$ such that
\[
m_1 = m_1^{\mathrm{ac}} + m_1^{\perp}, \quad m_1^{\mathrm{ac}} \ll m_2, \quad m_1^{\perp} \perp m_2.
\]
The Radon–Nikodym derivative ${\mathrm{d}m_1}/{\mathrm{d}m_2}$ is defined for the absolutely continuous part $m_1^{\mathrm{ac}}$.
 The recession constant $\varphi'_{\infty}$ is given by:
\begin{equation}
    \varphi'_{\infty} \coloneq \underset{s\to\infty}{\lim}\frac{\varphi(s)}{s} \in \mathbb{R} \cup \{ \infty \}.
\end{equation}
In the present work, we compute the divergence using the Kullback-Leibler (KL) entropy $\varphi_{\text{KL}}$ given by:
\begin{equation}
\varphi_{\mathrm{KL}}(s) =
\begin{cases}
s \log(s) - s + 1 & \text{for } s > 0, \\
1 & \text{for } s = 0, \\
+\infty & \text{otherwise.}
\end{cases}
\end{equation}
Note that when the total measures are equal (e.g., probability measures) and $m_1^{\perp}(\mathcal{X})=0$, $\mathcal{D}_{\varphi}(m_1 \mid m_2)$ with $\varphi_{\text{KL}}$ reduces to the familiar KL divergence from statistics.}

\section{Computation of the Unbalanced Optimal Transport Based Flow Field Dissimilarity}\label{appB}
As mentioned in Section~\ref{sec:OT}, the unbalanced optimal transport distance $d_\text{UOT}$ is defined as the infimum of the optimal transport cost $\mathcal{J}$ over all possible transport plans $\Gamma \in \mathcal{M}_+(\mathcal{X}_1 \times \mathcal{X}_2)$:
\begin{align}
\mathcal{J}(\Gamma) \coloneqq \int_{\mathcal{X}_1 \times \mathcal{X}_2} C(x_1,x_2) \, \mathrm{d}\Gamma(x_1,x_2)
+ \rho\left[\mathcal{D}_{\varphi_1}(P^{\mathcal{X}_1}_{\#}\Gamma \mid m_1) 
+ \mathcal{D}_{\varphi_2}(P^{\mathcal{X}_2}_{\#}\Gamma \mid m_2)\right],
\end{align}
\begin{equation}
d_\text{UOT}(m_1,m_2) \coloneqq \underset{\Gamma \in \mathcal{M}_+(\mathcal{X}_1 \times \mathcal{X}_2)}{\inf} \mathcal{J}(\Gamma).
\end{equation}
However, this optimization problem normally exhibits poor tractability. To compute the OT distance for histograms  
of dimension $N_h$, the computational cost scales at least in $\mathcal{O}(N_h^3\log(N_h))$, in the general case where no restrictions are placed upon the metric used to define the OT cost~\citep{pele2009fast, cuturi2013sinkhorn}. Consequently, rather than directly solving this constrained optimization problem, it is common to replace the non-negativity constraint on $\Gamma$ with an entropic regularization term to obtain an approximate solution. The entropic regularization changes the original linear programming problem to a strictly convex problem, which can be solved more efficiently using the Sinkhorn-Knopp matrix scaling algorithm~\citep{cuturi2013sinkhorn,chizat2016scaling}.

To define the entropically regularized problem, we consider the entropy of the transport plan. The negative entropy is given by
\begin{equation}
\mathcal{H}(\Gamma) = \int_{\mathcal{X}_1 \times \mathcal{X}_2} r(\log r-1)\diff(x_1,x_2),
\end{equation}
where the coupling $\Gamma$ is assumed to admit a density $r$ with respect to the reference Lebesgue product measure on $\mathcal{X}_1 \times \mathcal{X}_2$.

The entropically regularized unbalanced OT distance between measures $m_1$ and $m_2$ is the following convex optimization problem:
\begin{equation}
d_\text{UOT}^\varepsilon(m_1,m_2) = \min_{\Gamma \in \mathcal{M}(\mathcal{X}_1 \times \mathcal{X}_2)} (\mathcal{J}(\Gamma) + \varepsilon \mathcal{H}(\Gamma)),
\end{equation}
where $\varepsilon > 0$ is a small regularization constant. Here, the inclusion of the entropic regularization term $\mathcal{H}(\Gamma)$ follows the maximum-entropy principle and serves as a relaxation of the nonnegativity constraint on $\Gamma$~\citep{cuturi2013sinkhorn,chizat2016scaling}. 

It is important to note that the introduction of the entropic regularization results in a non-zero bias even when the input measures are identical~\citep{genevay2019sample,sejourne2019sinkhorn}. Because of this the following normalization is performed to obtain what is called the Sinkhorn divergence:
\begin{align}
S^\varepsilon(m_1,m_2) =& \ d_\text{UOT}^\varepsilon(m_1,m_2) - \frac{1}{2}[d_\text{UOT}^\varepsilon(m_1,m_1) + d_\text{UOT}^\varepsilon(m_2,m_2)] \nonumber\\ & + \frac{\varepsilon}{2}\left(\int_{\mathcal{X}_1}\diff m_1 - \int_{\mathcal{X}_2}\diff m_2 \right)^2,
\end{align}
which ensures that $S^\varepsilon(m_1,m_2) = 0$ when $m_1=m_2$. We note that since we choose the cost function $C$ to be the $\ell_2$ distance between two spatial points and only consider a compact spatial domain $\mathcal{X}$, we have that for all $\varepsilon>0$ the Sinkhorn divergence is positive definite and convex in each argument because the Euclidean distance is symmetric, $1$-Lipschitz with respect to the essential supremum norm, and $k_\varepsilon=e^{-C(\ \cdot \ , \ \cdot \ )/\varepsilon}$ is a positive universal kernel~\citep{sejourne2019sinkhorn}.

To compute the flow field dissimilarity for signed measures (such as vorticity), we decompose the flow field into positive and negative parts. Consider a single flow field variable over a given compact spatial domain represented by a function $V : \mathcal{X} \to \mathbb{R}$ assumed to be measurable. Using this function $V$, we define a signed measure with $\diff m = V(x,y)\diff x\diff y$ (here we use a two-dimensional domain $\mathcal{X}$ for notational simplicity, but the results are extensible to arbitrary spatial dimensions). This measure can be uniquely decomposed into two mutually singular positive measures $m^+$ and $m^-$, in this case given by:
\begin{align}
    \mathrm{d} m^+ &= V^+(x,y)\diff x\diff y, \\
    \mathrm{d} m^- &= V^-(x,y)\diff x\diff y,
\end{align}
where $V^+(x,y) = \max(\{V(x,y),0\})$ and $V^-(x,y) = -\min(\{V(x,y),0\})$ are the Radon-Nikodym derivatives of measures $m^+$ and $m^-$ with respect to the reference Lebesgue measure. These define the density of the positive and negative parts of the flow field at a given point in the spatial domain. We can interpret $m^+$ as the ``amount'' of $V^+$ contained in some area and the densities as the spatial distribution of $V^+$ (and analogously for $m^-$ and $V^-$).

We perform this decomposition for a field $V_1$, yielding measures $m_1^+$ and $m_1^-$ as well as for a second field $V_2$, yielding a second set of measures which we call $m_2^+$ and $m_2^-$. We define a dissimilarity by considering the Sinkhorn divergences between the positive and negative parts of each field separately:
\begin{equation}
d_\text{field}(V_1,V_2) = S^\varepsilon(m_1^+,m_2^+) + S^\varepsilon(m_1^-,m_2^-),
\end{equation} 
inspired by~\citet{mainini2012well,mainini2012description}.

Now if we have two sets of fields of $N$ variables, $\mathbf{V}_1=\{ V_1^{(i)} \}_{i=1}^{N}$ and $\mathbf{V}_2=\{ V_2^{(i)} \}_{i=1}^{N}$, we compute the OT based dissimilarity for each variable and consider a total dissimilarity using
\begin{equation}
\tilde{d}_{\text{field}}(\mathbf{V}_1,\mathbf{V}_2) = \bigoplus_{i=1}^N d_\text{field}(V_1^{(i)},V_2^{(i)}),
\end{equation}
where $\bigoplus$ denotes some permutation invariant aggregation operator, such as a (possibly weighted) sum, mean, or a vector $\ell_2$ norm. For this work, we take the $\ell_2$ norm of a vector consisting of the field dissimilarities in each variable.

\section{Autoencoder Model Details}\label{appC}

\begin{figure}
  \centerline{\includegraphics[width=0.97\textwidth]{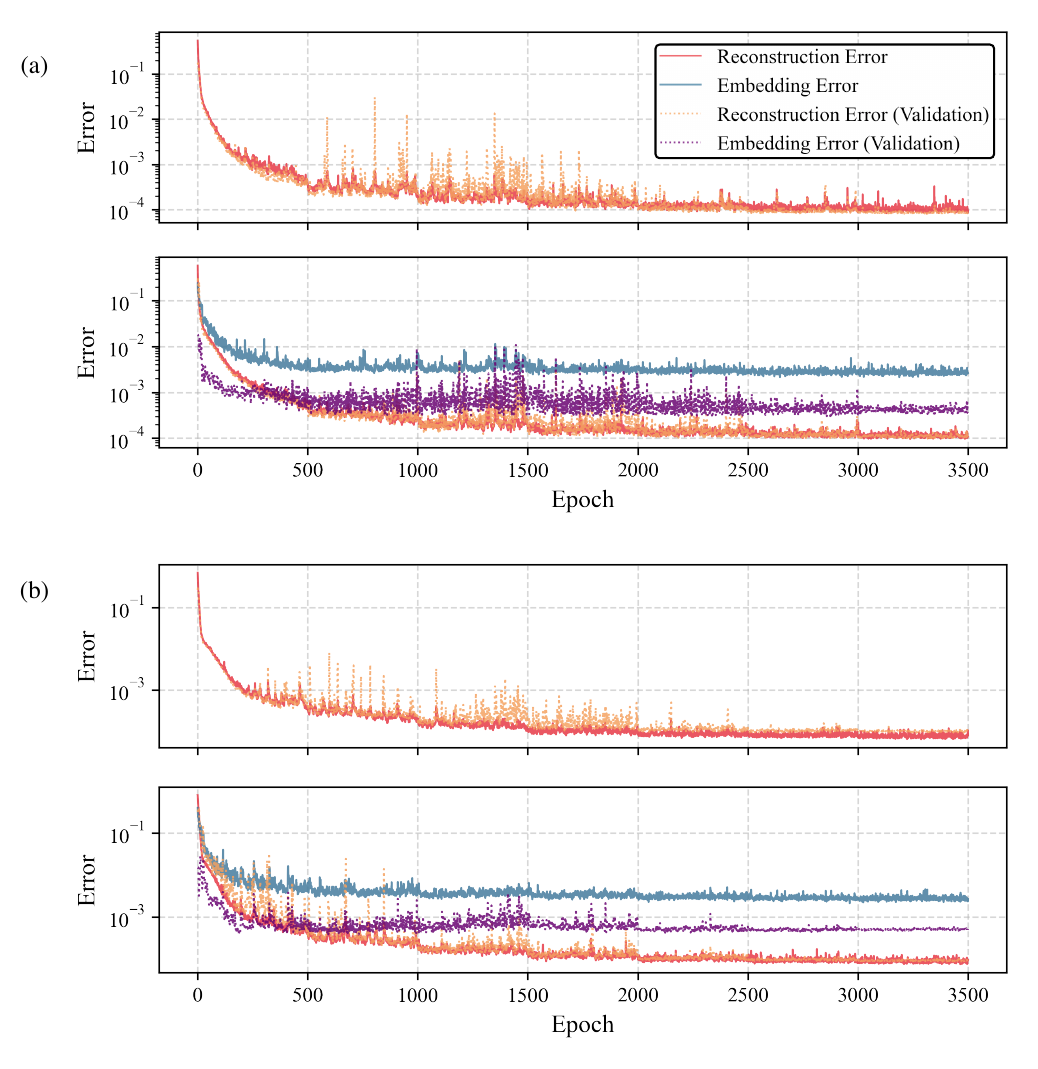}} 
  \caption{Plot of loss curves for (a) $\alpha=9^\circ$ and (b) $\alpha=6^\circ$ cases of both regular (top) and OT-based autoencoders (bottom).}
\label{fig:loss_curves}
\end{figure}

In the present analysis, we compare the learned latent spaces for two autoencoder architectures: a standard autoencoder that only uses reconstruction loss ($\mathcal{L}_1$ loss only) and the OT-based autoencoder ($\mathcal{L}_1$ and $\mathcal{L}_2$ losses). We note that both models can qualitatively reconstruct the flow fields at similar levels of accuracy. { We find that the inclusion of the OT-based embedding term does not have any noticeable impacts on the the training process in practice, as depicted by the example loss curves in Figure~\ref{fig:loss_curves}.}

\begin{figure}
  \centerline{\includegraphics[width=0.97\textwidth]{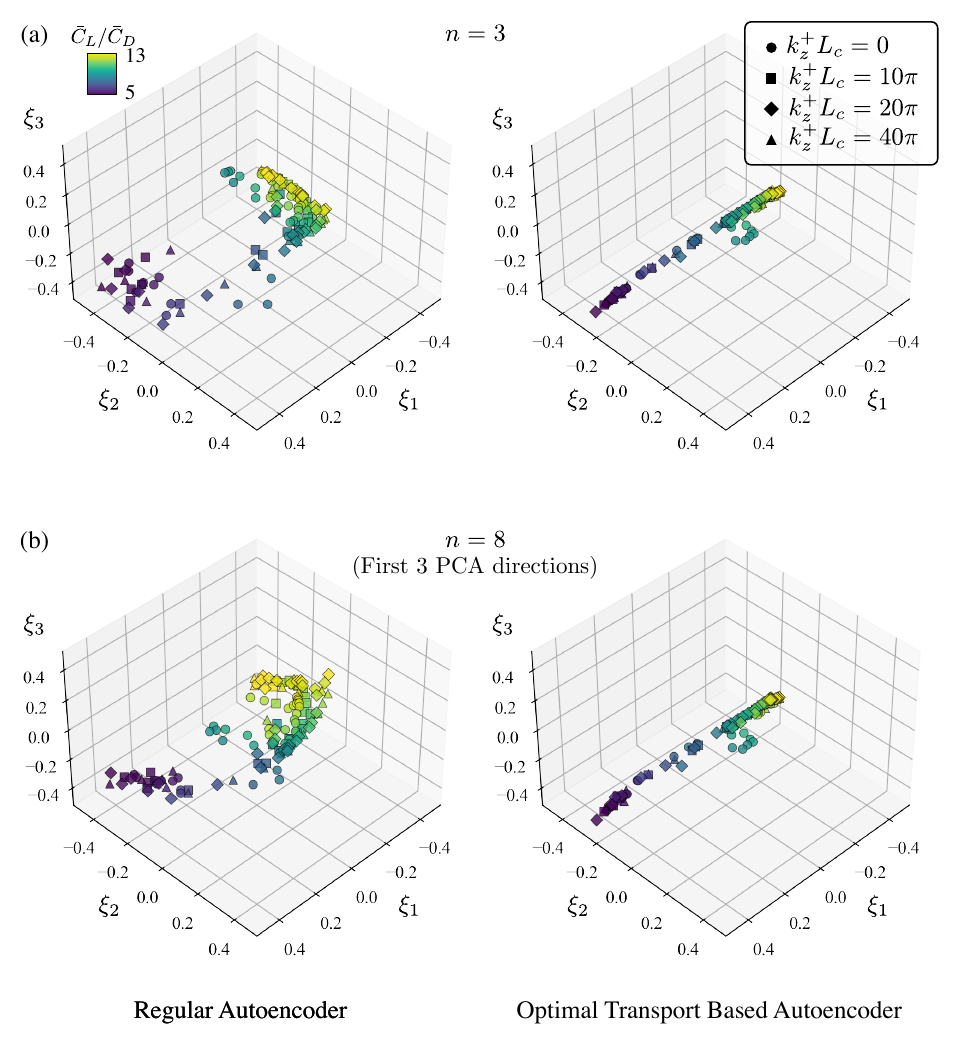}} 
  \caption{Plot of $\alpha=9^\circ$ latent space for both regular and OT-based autoencoder with (a) $n=3$ and (b) $n=8$ (showing the first 3 principal axes).}
\label{fig:latent_space_dim_aoa9}
\end{figure}

\begin{figure}
  \centerline{\includegraphics[width=0.97\textwidth]{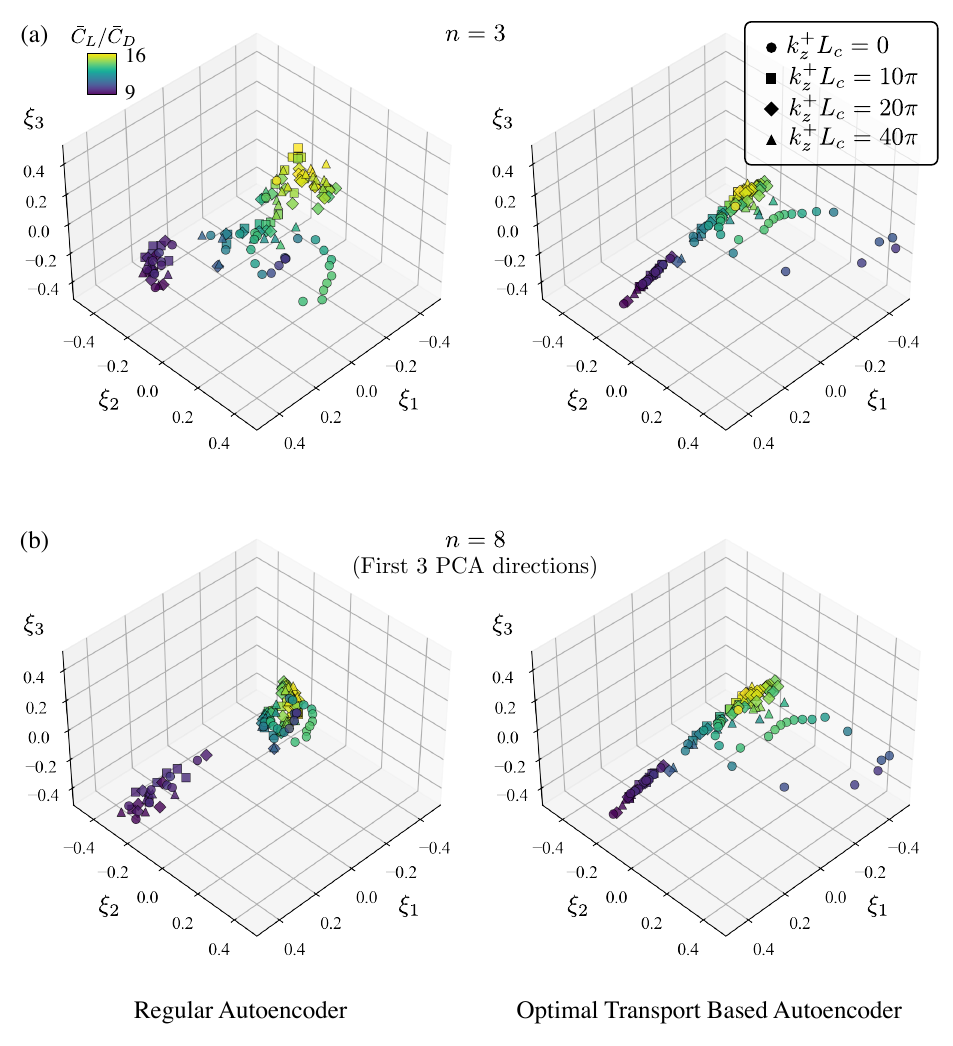}} 
  \caption{Plot of $\alpha=6^\circ$ latent space for both regular and OT-based autoencoder with (a) $n=3$ and (b) $n=8$ (showing the first 3 principal axes).}
\label{fig:latent_space_dim_aoa6}
\end{figure}

{ During the parameter study, we find that when increasing the latent space dimension the reduction in the loss slowed after $n=2$, prompting the use of a two-dimensional latent space. Additionally, we find that for higher-dimensional latent spaces, while the regular autoencoder produced mostly arbitrary representations of the latent geometry, the OT-based autoencoder consistently produced the same two-dimensional structure as seen in Figure~\ref{fig:latent_space_dim_aoa9} and~\ref{fig:latent_space_dim_aoa6} for $\alpha=9^\circ$ and $6^\circ$, respectively.}

{
\section{Pairwise Distance Comparison}\label{appD}

\begin{figure}
  \centerline{\includegraphics[width=0.9\textwidth]{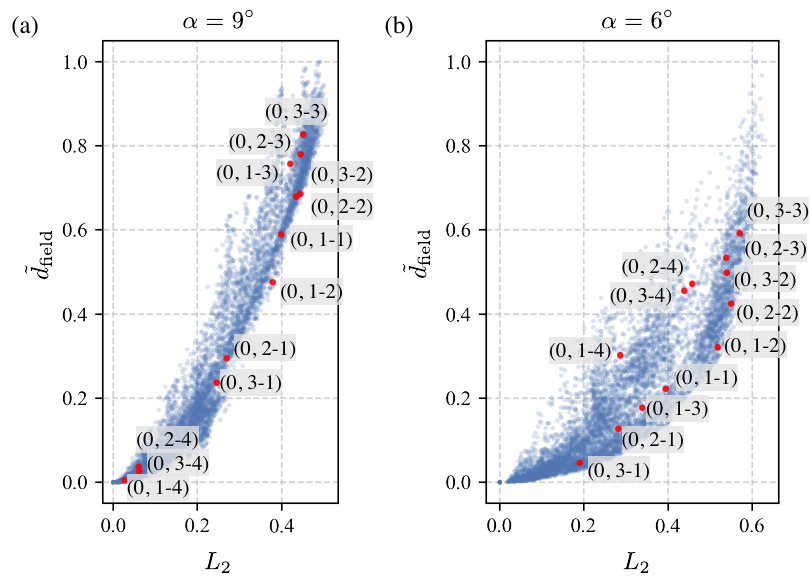}} 
  \caption{Plot of pairwise UOT-based distances ($\tilde{d}_\text{field}$) vs. pairwise $\L_2$ distances for discretized flow fields for (a) $\alpha=9^\circ$ and (b) $\alpha=6^\circ$. Both axes have been normalized by the maximum pairwise distance. Red points denote example pairwise distances from the baseline flow corresponding to the cases in Figure~\ref{fig:AoA9_Compare_Kz} and Figure~\ref{fig:AoA6_Compare_Kz}.}
\label{fig:distance_comparison}
\end{figure}

To illustrate the difference between the OT-based distance and the $L_2$ distance we plot pairwise UOT-based distances ($\tilde{d}_\text{field}$) vs. pairwise $L_2$ distances for discretized flow fields normalized by the maximum distance in the dataset in Figure~\ref{fig:distance_comparison}. Similar to the case with the previous example of the advecting and diffusing gaussian in Figure~\ref{fig:optimal_transport_advection}, while the distances appear correlated, the $L_2$ distance appears to saturate at around half the maximum value of $\tilde{d}_\text{field}$ after the separation bubble has been suppressed. This means that the UOT distance is capable of more effectively distinguishing cases as the separation bubble is suppressed. This intuitively manifests as a change in curvature, in which most cases are flattened along the $\xi_1$ axis in the OT-AE latent space.}

\FloatBarrier

\end{document}